# Optimal Number of Clusters by Measuring Similarity among Topographies for Spatio-temporal ERP Analysis

Reza Mahini[1+], Peng Xu[2+], Guoliang Chen[2], Yansong Li[3,4], Weiyan Ding[2], Lei Zhang[2], Nauman Khalid Qureshi[1], Asoke K. Nandi[5*], Fengyu Cong[1,6*]

[1] School of Biomedical Engineering, Faculty of Electronic and Electrical Engineering, Dalian University of Technology, China

[2] 215th Clinical Division, 406th Hospital of PLA, Dalian, China

[3] Department of Psychology, School of Social and Behavioral Sciences, Nanjing University, Nanjing, China

[4] The Research Center for Social and Behavioral Sciences of Jiangsu Province, Nanjing, China,

[5] Department of Electronic and Computer Engineering, Brunel University London, UK

[6] Faculty of Information Technology, University of Jyvaskyla, Finland

**Abstract**

Averaging amplitudes over consecutive time samples within a time-window is widely used to calculate the amplitude of an event-related potential (ERP) for cognitive neuroscience. Objective determination of the time-window is critical for determining the ERP component. Clustering on the spatio-temporal ERP data can obtain the time-window in which the consecutive time samples' topographies are expected to be highly similar in practice. However, there exists a challenging problem of determining an optimal number of clusters. Here, we develop a novel methodology to obtain the optimal number of clusters using consensus clustering on the spatio-temporal ERP data. Various clustering methods, namely, K-means, hierarchical clustering, fuzzy C-means, self-organizing map, and diffusion maps spectral clustering are combined in an ensemble clustering manner to find the most reliable clusters. When a range of numbers of clusters is applied on the spatio-temporal ERP dataset, the optimal number of clusters should correspond to the cluster of interest within which the average of correlation coefficients between topographies of every two-time sample in the time-window is the maximum for an ERP of interest. In our method, we consider fewer cluster maps for analyzing an optimal number of clusters for isolating the components of interest in the spatio-temporal ERP. The statistical comparison demonstrates that the present method outperforms other conventional approaches. This finding would be practically useful for discovering the optimal clustering in spatio-temporal ERP, especially when the cognitive knowledge about time-window is not clearly defined.

## 1. Introduction

Event-related potentials (ERPs) have been considered as a fundamental neuroimaging technique for cognitive neuroscience. The measuring mean peak amplitude of an ERP within a specific temporal interval, so-called time-window, undertakes a significant role in the statistical power analysis and is recommended for an ERP study (Luck 2014). The underlying assumption of this measurement is that the topographical maps over the

---

certain time-window are stable and can represent the quasi-stable synchronized network activation in the time-window (Lehmann 1990). Clustering of spatio-temporal ERP is used as a promising tool to achieve the time window (Brunet et al. 2011; Koenig et al. 2014; Murray et al. 2008). Evaluating the quality of the clustering in terms of the properties of an ERP has become more and more important.

So far, a general approach to evaluate the quality of clustering and the optimal number of clusters is to examine all the clusters in the result (Handy TC 2009; Koenig et al. 2014; Michel and Koenig 2017; Murray et al. 2008). However, this approach is not always appropriate for neurological interpretations of ERPs. The noteworthy issue in clustering is the trade-off between the number of clusters and their quality. If the number of clusters is low, only limited information can be extracted based on the explained variance. Therefore, the dataset will be highly compressed and not well segmented. On the other hand, by increasing the number of clusters, the explained variance will be high, but the dataset cannot be sufficiently compressed (Handy TC 2009; Murray et al. 2008). Hence, it is necessary to find an optimal number of clusters by balancing the trade-off between the quality of clustering and the extracted information for the ERP study.

In the literature of cluster analysis, there are several popular methods to obtain the optimal number of clusters for a given dataset. For instance, similarity within the objects (Kaufman and Rousseeuw 2009; Rousseeuw 1987), distance within the cluster objects and between the clusters (Dunn 1974), explained variance measurement (Goutte et al. 1999; Lleti et al. 2004) are widely used as classic methods for finding the optimal number of clusters. Furthermore, Milligan and Cooper (Milligan and Cooper 1985) considered thirty various indices to find the best clustering scheme from the results obtained by different quality measurement methods. Some of these indices are addressed in the *R* software package named NbClust (Charrad et al. 2014; Kassambara 2017). Additionally, a combination of some conventional methods (e.g. Elbow, Gap statistic, Silhouette) in the microstates analysis (Lehmann 1990; Mur et al. 2016) has also been commonly used to find an optimal number of clusters. The Information-based approaches (Jonnalagadda and Srinivasan 2009; Pelleg and Moore 2000; Sugar and James 2003) to analyze clustering quality such as; *X*-means method (Pelleg and Moore 2000) to optimize the Bayesian Information Criterion (BIC) or Akaike Information Criterion (AIC) measurement and recently, Gap statistic (Charrad et al. 2014) by evaluating within-group dispersion found more attention in recent decade. Together, the majority of the above-mentioned studies, in general, have focused on analyzing the quality of clustering by evaluating the tightness of clusters and distance between them.

Several studies have attempted to sketch an ERP model by assigning individual ERP maps to the specially defined clusters obtained from clustering the group grand-averaged data (Brunet et al. 2011; Koenig et al. 2014; Murray et al. 2008; Pascualmarqui et al. 1995; Pourtois et al. 2008; Ruggeri et al. 2019; Wackermann et al. 1993). Limited clustering methods such as; K-means or hierarchical clustering with cross-validation method (Pascualmarqui et al. 1995) for an optimal number of microstate maps detection have been used in most of those studies. Other authors considered the modified cross-validation criterion (Kawamoto and Kabashima 2017); however, this method often fails to find the minimum value when the high-density montages. To deal with this problem, Tibshirani et al (Tibshirani et al. 2001) have developed Krzanowski–Lai (KL) criterion based on agglomerative hierarchical clustering (AAHC) dispersion. More advanced methodologies have employed a hybrid criterion from several optimization criteria (Custo et al. 2017) by running the clustering method for many times to select a suitable number of clusters (microstate maps). These approaches use meta-criterion based on calculating the ranking for several criteria taken from different methodologies (Michel and Koenig 2017). Additionally, a cluster validity index, named S-Dbw, based on clusters' compactness (i.e., intra-cluster variance), and density between clusters (i.e., inter-cluster density) have been utilized for the selection of optimal input parameter values for a clustering algorithm (Halkidi and Vazirgiannis 2001; Oostenveld et al. 2011). Overall, these studies have been conducted to explore the ERP microstate model based on whole spatio-temporal

ERP data analysis using a single clustering method. However, a reliable and objective method to find an optimal number of clusters for spatio-temporal ERP is a gap in this research field.

For filling that gap, we propose a new approach to exploring an optimal number of clusters using the cluster ensemble method. Indeed, the rationale of the proposed method includes three important points: Firstly, in an ERP dataset, several ERP components are inevitably generated, however, a few of them are targeted ERP components which are more probably elicited if the ERP experiment is run again. Moreover, those targeted ERP components are more probably elicited among multiple subjects. This allows us to analyze a certain ERP when the clustering is applied to the ERP data. Secondly, for an interesting ERP component, clustering the ERP dataset into different numbers of clusters may affect its analysis. This is because of two reasons: The first reason is that the ERP component can be associated with the certain brain activity which has its own topography in the spatial (i.e., topographic) domain and has its own starting time-point and the ending time-point in the time domain. The other reason is, the inappropriate number of clusters used for the clustering may result in that the separation of one true cluster into two or more clusters in practice. Therefore, finding the appropriate number of clusters used for the clustering is very critical. Finally, the ideal number of clusters used for the clustering can result in the perfect cluster of interest in theory. For the perfect cluster, the topographies of different time samples in a time-window will be identical since the cluster represents an ERP component of the certain brain activity. Therefore, the correlation coefficient of the topographies between any two time points in the time window found by the clustering should be 1 in theory and be closer to 1 in practice.

For finding an optimal number of clusters, the proposed method investigates the mean inner-similarity of the selected time-windows from the clustering options (e.g. from 2 to 15 clusters in this study based on the past experiences, which can be increased when it is needed) by the procedures many times. The inner-similarity of a cluster can be defined as the mean of correlation coefficients between topographical maps within any two time samples of the found time window by the clustering.

The major contributions of this study can be summarized as follows:

- Clustering of spatio-temporal ERP via consensus clustering;
- Identifying an optimal time-window by considering the component of interest;
- Selecting the most reliable number of clusters;

Therefore, we have implemented five standard clustering algorithms for spatio-temporal ERP and applied them on two ERP datasets namely, simulated ERP data and real ERP data which is about prospective memory (PM) from our group's previous publication (Chen et al. 2015).

**2. Materials and Methods**

In this section, after a brief background about consensus clustering, we provide more detail about the proposed method. We also discuss briefly the two types of ERP data (i.e. simulated ERP and real data) which have been used for the evaluation of our method performance. Meanwhile, the statistical power analysis method and studied factors for evaluation of our method are discussed.

**2.1 Consensus Clustering**

In theory, combining different clustering results, also known as ensemble clustering or consensus clustering has received a lot of attention in different research areas (Abu-Jamous et al. 2013; Basel Abu-Jamous 2015; Fred and Jain 2005; Meila 2007). The rationale of consensus clustering method is to solve the problem of inconsistency of stochastic clustering algorithms or clustering with different parameters. Furthermore, consensus clustering technique has been used successfully in different biological data processing (Abu-Jamous et al. 2013; Abu-Jamous et al. 2015; Liu et al. 2015; Mahini et al. 2017; Monti et al. 2003) with vary ensemble functions. It has also been utilized in the human brain research area i.e. functional magnetic resonance imaging (fMRI) data processing (Liu et al. 2017a; Liu et al. 2017b). According to Vega et al. study (Vega-Pons and Ruiz-Shulcloper 2011) the most important advantages of cluster ensemble are; **i)** *Robustness*, to achieve better averaged performance than single clustering method, **ii)** *Consistency*, the final ensemble results are correlated with single algorithm results, **iii)** *Novelty*, finding solution unattainable by single clustering and, **iv)** *Stability*, results are less sensitive to noise and outliers.

Let us define the following uniform notation as the basic background of consensus clustering. $X = \{x_1, x_2, ..., x_n\}$ is a set of objects/observations (i.e. time samples in this study), each object $x_i$, $i = 1, ..., n$ is described by a set of m-dimensional features space recorded voltage from the scalp, $F = \{f_1, f_2, ..., f_m\}$. Where each observation can be represented as a vector such as $x_i = \{d_{i1}, d_{i2}, ..., d_{im}\}$. In order to group *n* data objects into *K* clusters, we employ cluster ensemble by combining *R* different clustering results (i.e. $L = \{L_1, L_2, ..., L_R\}$). Where each labeling can be demonstrated by $L_r = \{C_1^r, C_2^r, ..., C_K^r\}$, $r = 1, ..., R$ is a clustering result of *X* with *K* clusters. Therefore, $C_k^r$ is the *kth* cluster, $k = 1, ..., K$ of *rth* clustering method. The goal of clustering ensemble method is to find a consensus labeling such as $L^*$ which could better represent the properties of each labeling in *L* in terms of specificity and coverage of the information in the dataset.

$$X = \begin{pmatrix} d_{11} & d_{12} & \cdots & d_{1m} \\ d_{21} & d_{22} & \cdots & d_{2m} \\ \vdots & \vdots & \ddots & \vdots \\ d_{n1} & d_{n2} & \cdots & d_{nm} \end{pmatrix} \quad (1)$$

Mathematically, we can define $L^*$ as follows:

$$L^* = argmax_{L \in \mathbb{L}} \sum_{r=1}^{R} \Gamma(L, L_r) \quad (2)$$

also

$$L^* = argmin_{L \in \mathbb{L}} \sum_{r=1}^{R} \mathcal{M}(L, L_r) \quad (3)$$

where $\Gamma$ is a similarity measurement and $\mathcal{M}$ is a dissimilarity measurement, which can measure mutual information between a set of *R* labelings. From the Equation 2 and 3, the cluster ensemble $L^*$ is an optimally combined clustering with maximum similarity to other clusterings and minimum dissimilarity to them, we call it clustering objective function.

**Table 1** The illustration of cluster ensemble problem with 5 clustering methods, K=3, number of time samples in data n=6: Original labelings (left) and the hypergraph representation with 15 hyperedges (right) Each cluster map is transformed into a hyperedge.

|  | $l_1$ | $l_2$ | $l_3$ | $l_4$ | $l_5$ |  | $H^{(1)}$ | | | $H^{(2)}$ | | | $H^{(3)}$ | | | $H^{(4)}$ | | | $H^{(5)}$ | | |
|---|---|---|---|---|---|---|---|---|---|---|---|---|---|---|---|---|---|---|---|---|---|
|  |  |  |  |  |  |  | $h_1$ | $h_2$ | $h_3$ | $h_4$ | $h_5$ | $h_6$ | $h_7$ | $h_8$ | $h_9$ | $h_{10}$ | $h_{11}$ | $h_{12}$ | $h_{13}$ | $h_{14}$ | $h_{15}$ |
| $x_1$ | 1 | 1 | 1 | 1 | 2 |  | 1 | 0 | 0 | 1 | 0 | 0 | 1 | 0 | 0 | 1 | 0 | 0 | 0 | 1 | 0 |
| $x_2$ | 1 | 2 | 1 | 2 | 1 |  | 1 | 0 | 0 | 0 | 1 | 0 | 1 | 0 | 0 | 0 | 1 | 0 | 1 | 0 | 0 |
| $x_3$ | 2 | 2 | 1 | 1 | 2 | ⇔ | 0 | 1 | 0 | 0 | 1 | 0 | 1 | 0 | 0 | 1 | 0 | 0 | 0 | 1 | 0 |
| $x_4$ | 1 | 3 | 3 | 3 | 2 |  | 1 | 0 | 0 | 0 | 0 | 1 | 0 | 0 | 1 | 0 | 0 | 1 | 0 | 1 | 0 |
| $x_5$ | 3 | 3 | 3 | 3 | 3 |  | 0 | 0 | 1 | 0 | 0 | 1 | 0 | 0 | 1 | 0 | 0 | 1 | 0 | 0 | 1 |
| $x_6$ | 3 | 3 | 2 | 2 | 3 |  | 0 | 0 | 1 | 0 | 0 | 1 | 0 | 1 | 0 | 0 | 1 | 0 | 0 | 0 | 1 |

Among the variety of consensus functions, we investigate the cluster-based similarity partitioning algorithm (CSPA) which is based on pairwise similarity measurement between partitions (Karypis and Kumar 1998; Nguyen and Caruana 2007). This method reclusters the samples yielding ensemble clustering. The hypergraph-based cluster ensemble is useful when multiple clustering algorithms with different labeling results were used for clustering (Strehl and Ghosh 2002). Therefore, we implement the CSPA method in MATLAB platform for the spatio-temporal ERP clustering with various clusterings. As a result, the consensus function (i.e. CSPA) transforms the cluster labelings obtained from five different clustering methods into a suitable hypergraph like $H$ representation. By creating hyperedges such as $H^{(r)}$ of the hypergraph made by multiple labelings, a new dataset is created define as:

$$H = H^{(1,\dots,R)} = \left(H^{(1)} \dots H^{(R)}\right) \qquad (4)$$

where the number of hyperedges $HE$ denote by,

$$HE = \sum_{r=1}^{R} K^{(r)} \qquad (5)$$

Thus, for each labeling such as $L_r$, a binary membership matrix $H^{(r)}$ with a column of the cluster (called hyperedge) is defined. Hence, $H$ defines the adjacency matrix of a hypergraph with $n$ objects and $HE$ denotes the hyperedges which $K^{(r)}$ is the number of clusters in $rth$ method. Therefore, we have mapped each individual clustering to a hyperedge and all clusterings to hypergraph. The Table 1 demonstrates an example of 6 observations and 5 labeling results for cluster ensemble task with 3 cluster maps. Hence, we can obtain a $n \times n$ binary similarity matrix for each clustering. Thus, entry-wise averaging of $R$ clusterings yields an overall similarity matrix $S$ with a high granular resolution. The entries of $S$ denote the fraction of clusterings in which two objects are in the same cluster, and can be computed in one sparse matrix multiplication,

$$S = \frac{1}{R} HH' \qquad (6)$$

this is named as the cluster-based similarity matrix. Next, we recluster the combined similarity matrix (i.e. overall similarity matrix S) with hierarchical clustering as the final step for consensus clustering.

**2.2 Proposed Method**

The proposed method includes two procedures (i.e. *Procedure 1* and *Procedure 2)*. The main procedure (i.e. *Procedure 1*) includes four main steps: **i**) *Dataset concatenating*, **ii**) *Generation phase and consensus clustering*, **iii**) *Time-window selection* and, **iv**) *Selecting the optimal number of clusters*. We illustrate those steps in Fig.1, and in Procedure 1 in more details.

*a) Concatenating ERP Dataset*

The underlying assumption for configuring the concatenated dataset is that a brain activity respects each stimulus/condition which comes from a similar brain response from all the subjects in practice. However, it is accepted that many criteria such as age, gender, sleep, etc. could influence the response to the individual subject. To acquire more effective cluster learning, we provide a concatenated ERP dataset (Micah et al. 2009) for each subject by linked-combining datasets of all conditions. Then by grand-averaging the subjects' datasets across each group and linked-combining the datasets from multiple-group, a larger dataset is provided for further processing. This procedure has been shown in Fig. 1b.

*b) Generation Phase and Consensus Clustering*

To generate partitioning results in $K$ partitions, multiple clustering algorithms are applied over the concatenated data acquired in the previous step. We employ five clustering algorithms, namely, *K*-means (Yu et al. 2012), hierarchical clustering (Tan 2006), fuzzy C-means (FCM) (Bezdek 1981), self-organizing maps (SOM) (Kohonen 1990) and diffusion map spectral clustering (Sipola et al. 2013a). The proposed method generates many partitions from various clustering methods regarding clusters options (e.g. 2 to 15 clusters) to feed consensus function. The consensus clustering generates the mutual contribution within the clusterings in an ensemble partitioning from generated partitions using the concept of hypergraph partitioning to subgraphs.

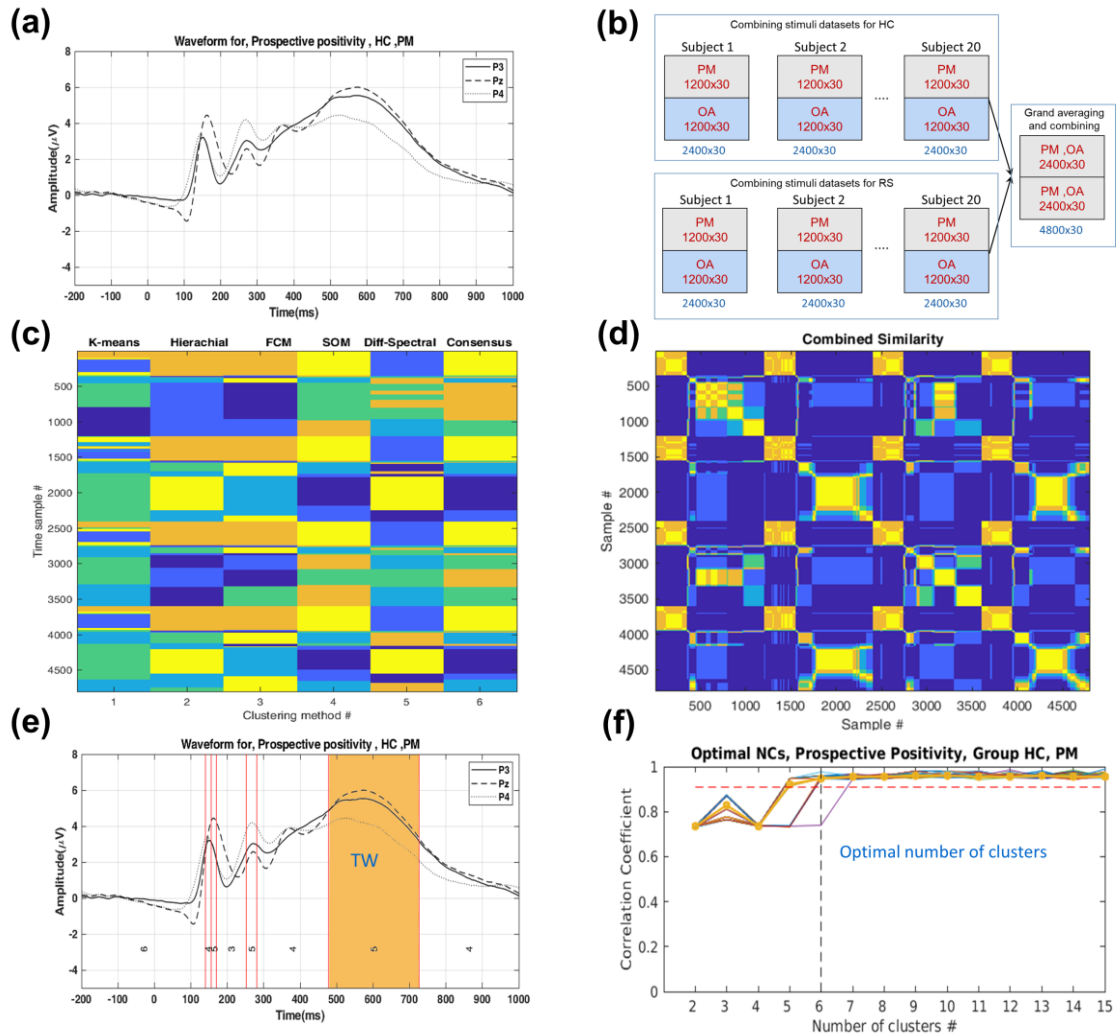

**Fig. 1** Illustration of the proposed approach: **a.** A grand-averaged ERP waveform for studying prospective memory (PM) paradigm. **b.** Providing restructured ERP data across groups/conditions. **c.** Illustration the result of different clustering methods for restructured ERP data. **d.** Combined similarity matrix used in consensus function generated by averaging similarity matrixes of 5 clustering methods. **e.** The partitioned ERP waveform and selected most proper time-window (TW) in order to highest inner-similarity and better coverage with predefined measurement interval. **f.** Recording the time-window inner-similarities for each clustering option (i.e., 2 to 15 maps). The new structured ERP data is used in the cluster analysis task for the spatio-temporal domain. A new criterion called inner-similarity (see the text) recorded for selected TWs. The mean inner-similarity of selected TWs for iterative using consensus clustering (orange line) reveals 6 reliable cluster maps for this dataset. PM= prospective positivity, HC=healthy controlled, RS=Remitted schizophrenia OA=ongoing activity, NC=number of clusters.

```
Procedure 1: Optimal Number of Clusters Algorithm

Input: ERP data, Components of interest
Output: Optimal number of clusters
Procedure {
   Step1. Concatenating dataset, preparing dataset and grand average across the subjects;
   For 100 iterations
      For each Component of interest
         For clustering options (Number of clusters=2 to 15)
            Step2. Generation, providing clusterings by clustering algorithms; and then consensus
                   clustering, applying consensus clustering to find ensemble of clusterings;
            Step3. Selecting best time-window and saving its inner-similarity;
         End of For (Number of clusters).
      End of For (component of interest)
   End of iterations
   Step4. Selecting the optimal number of clusters (i.e., from mean of inner-similarities within iterations);
} End of Procedure
```

*c) Time-window Selection*

The time-window selection procedure provides the most suitable time-window for the component of interest by employing inner-similarity as a criterion for evaluating the quality of candidate cluster maps among all cluster maps. The candidate cluster maps (clusters of interest) can be defined as the clusters in the interested measurement interval roughly defined by the experimenter. More details and procedure steps are explained as the steps in Procedure 2. To clarify the mechanism of the time-window selection procedure, some definitions are noted.

```
Procedure 2: Time-Window Selection Algorithm

Input: Clustering result, Component of interest
Output: Time-window
Procedure {
Step1. Finding the candidate microstate maps;
        For selected maps
             Step2. Calculating inner-similarity and ranking the maps;
             Step3. Selecting high ranked maps;
             Step4. Selecting the highly correlated and overlapped maps regarding the component of
                    interest.
        End of Selecting the best time-window (End of For)
} End of Procedure
```

Technically, the spatial correlation could be defined as the correlation coefficient between two conditions at the same time point, or between two different time points (Murray et al. 2008). It can be noted by this equation,

$$\bar{u} = \frac{1}{E} \cdot \sum_{e=1}^{E} U_e \tag{7}$$

where

$$u_e = U_e - \bar{u} \tag{8}$$

where $E$ is the number of electrodes in the montage, $\bar{u}$ is the mean value of all $U_e$'s (for a given condition, at a given time point $t$) $u_e$ is the average-referenced potential of the *eth* electrode (for a given condition, at a given time point $t$ and $v_e$) is the measured potential of the $eth$ electrode, either from another condition $V$, or from the same condition $U$, but at a different time point $t'$.

$$Co_{u,v} = \frac{\sum_{e=1}^{E} u_e . v_e}{\|u\| . \|v\|} \tag{9}$$

where

$$\|u\| = \sqrt{\sum_{e=1}^{E} u_e^2}, \|v\| = \sqrt{\sum_{e=1}^{E} v_e^2} \tag{10}$$

The correlation coefficient between samples in a time-window can be denoted by

$$C = Corr(TS) \tag{11}$$

where $TS$ is the matrix (i.e. size of $NS$ samples and $E$ electrodes) for time samples and voltage vectors in a microstate map. Note that $C$ is Pearson correlation coefficient matrix form $NS$ time samples. We define a distance matrix for the elements in the correlation matrix as

$$D_{ns} = d(C_{ns,j}, C_{ns,ns})$$
$$D_{zns} = arctanh(D_{ns}) \tag{12}$$

where $D$ is the distance matrix size of $NS \times (NS - 1)$ which $ns = 1, ..., NS$ denotes the rows number and $j = 1,2, ... NS$, $j \neq ns$ and $C_{ns,ns} = 1$ (i.e. self correlation). Therefore, we calculate Fisher z-transform (Fisher 1921) for each vector $D_{ns}$ named $D_{zns}$. Next, we calculate the average for $D_{zns}$ we named it $D_{avg}$ after inverse z-transform. Then for each candidate cluster map standard deviation of $D_{avg}$ is evaluated.

$$D_{av} = \frac{\sum_{ns=1}^{NS} D_{zns}}{NS} \tag{13}$$

$$D_{avg} = tanh(D_{av})$$
$$STD_{TS} = STD(D_{avg}) \tag{14}$$

Therefore, after calculating this criterion for all candidate cluster maps $STD_{TS}$ is calculated. Two minimum values among candidates are selected for selecting time-window. The average of the correlation coefficient for the selected cluster map is considered as inner-similarity by,

$$CZ = arctanh(C)$$
$$InSim = tanh\left(\frac{\sum_{a,b=1}^{NS} Corr(CZ_{a,b})}{NS}\right) \tag{15}$$

where $C$ is the Pearson cross-correlation coefficient, $a$ and $b$ are two different time samples in the cluster map and $InSim$ demonstrates the inner-similarity of the cluster map.

Noteworthy, one may find a cluster map with high inner-similarity, but low coverage ratio for the component of interest, which is not suitable to be selected as time-window. Hence, the time-window selection algorithm selects two high ranked cluster maps among the candidates. This can give a chance to probable second cluster map candidate to be selected as time-window. Thus, we also measure the overlap ratio with the component of interests for time-window candidates. Therefore, the higher overlapped cluster map with high inner-similarity has more priority to be selected as the time-window. For example, if the time-window selection algorithm detected three cluster map candidates in the interested temporal interval, with inner-similarities of 0.82, 0.93 and 0.65 respectively, then the time-window selection method selects the two highest rank cluster maps as the best time-window candidates (i.e. 0.93 and 0.82). Consequently, by calculating the overlap between the interested interval and those cluster maps, the highest-ranked and highly overlapped cluster map is selected as the best time-window.

### d) Optimal Number of Clusters

The proposed method runs both procedures (i.e. Procedure1 and Procedure2) in 100 times and measures inner-similarities of the selected time-windows recorded from each clustering option. We concerned the satisfactory inner-similarity of 0.95 as a threshold for the satisfaction level of selecting the optimal number of clusters. The proposed method explores an optimal number of clusters from the mean inner-similarities behavior calculated from the inner-similarity of the selected time-windows in 100 runs by investigating two key criteria, Inner-similarity qualities, and stability. If the inner-similarity meets the satisfactory level, then the method will explore for a stable point as the optimal choice. The stable point can be defined as a number of clusters choice which, the difference between inner-similarities before and after it does not higher than 0.03 (i.e. the maximum change in inner-similarities between two sequential time-windows). In other words, if the inner-similarity of selected time-window converges to a reasonably satisfactory and stability level compared to other time-windows, it will be determined as an optimal number of clusters for further analysis.

## 2.3 ERP Data

- *Simulated ERP data*

In order to validate the proposed methodology, we applied the proposed approach on simulated ERP data. The simulated data were conducted using Berg's Dipo (Berg 2006) simulator. We defined 6 components, namely, P1, N1, P2, N2, P3, and N4 and two tasks which simply named St1 and St2 include a group of 20 subjects. We used to select the scalp with 65 electrodes for representing the recorded power and topography analysis. Each epoch starts from 100 ms pre-stimulus to 600ms post-stimulus with 214 Hz sampling rate. The averaged reference method has been used for the referencing. The defined components' topography maps and corresponding waveforms have been demonstrated in Fig. 2. Two interesting components, namely, N2 which refers to the maximum negative peak in 183 to 278 ms and P3 which refers to the positive response in 231 to 350 ms were selected for more detail analyzing. Since we were curious about the performance of the proposed method against the noisy signal, signal to noise was manipulated using MATLAB function *awgn* (i.e., add white Gaussian noise) to add noise (i.e., 20 dB) to signal power measured for each simulated dataset as a whole. The purpose of adding this noise is generating data similar to the real ERP data. The data are supposed to be preprocessed and the components are pre-defined. The electrode sites, for measuring statistical amplitude power difference are P2/P6/PO4 for N2 and CP2/CPz/Cz for P3.

- *Real ERP data*

We also applied the proposed method in real (PM experiment) (Chen et al. 2015) ERP data for demonstrating the most essential features of the proposed approach. Following the prior study (Chen et al. 2015), the experiment data includes 20 symptomatically remitted patients with schizophrenia (RS) and 20 healthy controlled (HC). Two tasks, namely, PM and ongoing were investigated following the previous study. Electroencephalogram (EEG) data was recorded (SynAmps amplifier, NeuroScan) with a quick cap, carrying 32 Ag/AgCl electrodes placed at standard locations covering the whole scalp (the extended international 10–20 system). The electrophysiological data was continuously recorded with a bandwidth of 0.05–100 Hz and sampled at a rate of 1000 Hz. The data initially was preprocessed by EEGLAB (Delorme and Makeig 2004) and first re-referenced to linked mastoid (A1 and A2). An independent component analysis (ICA)-based artifact correction was achieved using the ICA function of EEGLAB. Independent components with topographies representing saccades blinks and the heart rate artifact was thus removed according to the published guidelines (Jung et al. 2000). The resultant EEG data were then epoched from 200ms pre-stimulus to 1000 ms post-stimulus and digitally low pass filtered by 30 Hz (24 dB/octave). The 200 ms pre-stimulus period was used for baseline correction. In order to remove movement artifacts, epochs were rejected when fluctuations in potential values exceeded ±75μV at any channel except the EOG channel. The ERPs evoked by PM cue trials and ongoing activity trials were calculated by averaging individual artifact-free trials in each participant. The average across all channels was used as a reference. Two target ERP components, namely, N300 and prospective positivity components were studied in this research similar to (Chen et al. 2015).

## 2.4 Statistical Analysis

For evaluating the statistical performance of the new methodology, we measured the mean amplitude over selected time-window (i.e., as the measurement interval) for each condition/group and sensor sites for each study (i.e., simulated, and real ERP). We applied a 2 x 3 mixed repeated-measures ANOVA with two within-subject factors: Electrode (sensor sites) for analyzing N2 component, namely, P2/P6/PO4 and CP2/CPz/Cz for P3 component and *Task* which we named the simply, 'St1' and 'St2' for simulated ERP data. The measurement time-windows for conditions were calculated via the time-window selection procedure. This was set up for the mean amplitude of N2 and P3 separately clustered and calculated on selected electrodes according to the components of interest. The standard analysis was carried out to determine whether these effects of the noted factors for each study were statistically significant.

Concerning statistical power analysis on real ERP, statistical analyses carried out via utilizing a Greenhouse–Geisser corrected repeated measures ANOVA (i.e. mixed 2 x 2 x 3) with addition a between-subject factor: *Group* (RS and HC) and two within-subject factors: *Task* (PM and Ongoing) and *Electrode* (sensor sites: O1/Oz/O2 for N300 and P3/Pz/P4 for Prospective positivity). This was set up for the mean amplitude of N300 and Prospective positivity measured in the selected time-window. Therefore, the ERP statistical analysis involved analysis of two PM ERP components: N300 and prospective positivity. The N300 referred to the maximum negative voltage over the occipital region between 190 and 400 ms and the prospective positivity represented the maximum positive voltage over the parietal region between 400 and 1000 ms. Despite that, we use the measurement window calculated via time-window selection procedure for the optimal number of clusters for each condition/group. It is worth to mention, the selection of electrodes characterizing N300 and prospective positivity was based on prior ERP findings (Chen et al. 2015). Specifically, the amplitude of N300 over the occipital region (electrodes: O1/Oz/O2) was quantified and prospective positivity over the parietal region (electrodes: P3/Pz/P4) was measured. Statistical comparisons were made at *p*-values of $p < 0.05$.

## 3. Results

We would like to rely on the proposed method to demonstrate the optimal number of clusters selection for simulated and real ERP data. Then we have provided several analyses to demonstrate the repeatability and the performance of the proposed approach in 100 runs for both ERP data. We have first established the inner-similarity results for the selected time-windows in the various numbers of clusters (i.e. from 2 to 15 clusters). Next, we have revealed that how the proposed method can obtain the optimal number of clusters by processing the information obtained in many times recording the inner-similarities (i.e., 100 runs). Furthermore, we have also compared the results by conventional methods with the results of the proposed method in the selection of the optimal number of clusters. Later, the statistical power analysis results have been provided via measuring mean amplitude on selected time-windows. Finally, we also have presented the mutual contribution of individual clustering methods in consensus clustering and the quality of the results.

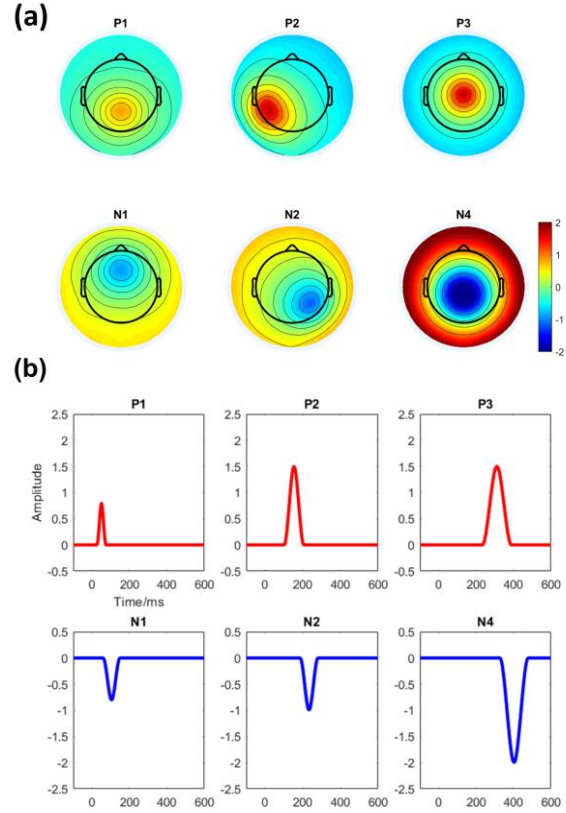

**Fig.2** Illustration of topography maps and waveforms for the defined components, namely, P1, N1, P2, N2, P3, and N4 in simulated ERP data. **a.** The topography maps of the components. **b.** Represents the components' defined waveforms.

### 3.1 Results for Simulated ERP Data

Before reporting the results of applying the proposed methodology in simulated ERP data, a brief discussion about the data properties is necessary. First, to show all the defined components in data, we have applied consensus clustering on the grand-averaged simulated ERP data (concatenated dataset) with 9 clusters. One may argue about selecting 9 clusters, the reason to select 9 clusters is, we can clearly detect all the defined 6 components by visual inspection (i.e., considering each cluster's topography and its temporal property), while by using 6 or 7 clusters, all the components are not distinctly detectable in the clustering results and only stronger components are elicited. We are able to select the number of clusters with 10 or 11 clusters; however, these components are not distinctly noticeable, because of obtaining many cluster maps which are not suitable to consider as a component. Fig. 3. reveals the topography maps and the waveform selected electrode sites, namely, Pz, Fz, CP5, P6, Cz, and CPz. The cluster maps; 6, 8 and 3 are corresponding to P1, P2, and P3 components, and the cluster maps; 5, 2 and 7 are corresponding to N1, N2, and N4 components, respectively. The cluster 9 does not represent any ERP component and cluster 4 is not a reliable cluster to be considered as a component. Therefore, 6 components were truly elicited by the proposed consensus clustering method. The followings are the proposed method results in simulated ERP based on different criteria.

- *Inner-similarity Analysis and the Optimal Number of Clusters Selection in Simulated ERP Data*

Fig. 4 exhibits inner-similarity of selected time-windows in 100 runs and a various number of clusters options (i.e. 2 to 15 clusters). Our strategy for selecting the optimal number of cluster maps is to select a reasonable number of clusters from the mean of inner-similarity results when the behavior of mean inner-similarity meets at a stable point and high inner-similarity threshold. In one extreme the optimal number of clusters for N2 in both stimuli 1 and 2 meet at 6 cluster maps. While, in another extreme, they are 5 and 4 cluster maps for the stimulus 1 and 2 of P3 component, respectively. Overall, for all the conditions/tasks, 5 cluster maps seem to be a fairly reasonable optimal choice among all other cluster maps. Therefore, the proposed method selects 5 cluster maps by processing the average of inner-similarities. This entails that the optimal number of clusters can be chosen according to the number of clusters with a high satisfactory threshold (i.e. 0.95) and stability of inner-similarities. Ultimately, we have operated the results of clustering in the optimal number of clusters for further analysis.

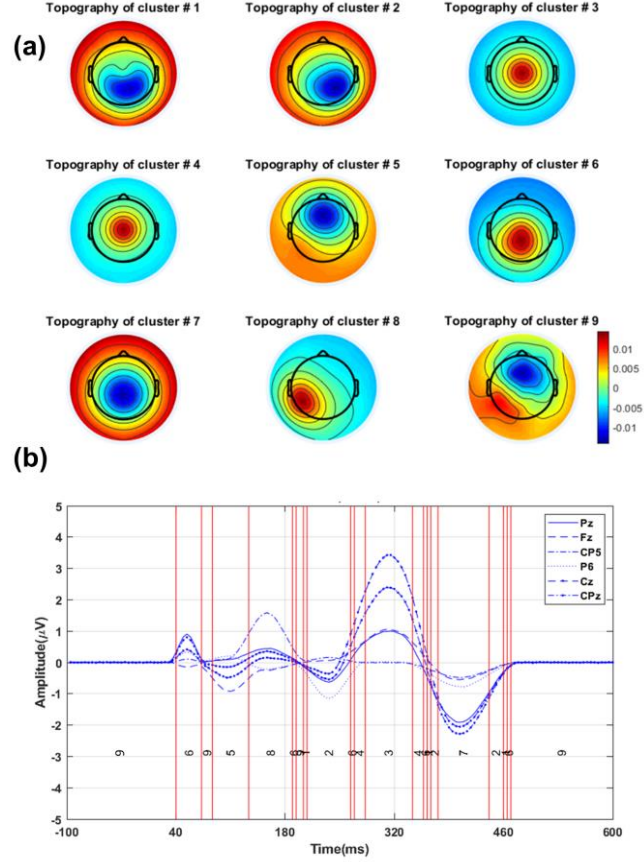

**Fig. 3** Consensus clustering results on grand-averaged ERP for extracting all 6 components. By using 9 clusters, all 6 components are extractable in the grand-averaged ERP data. **a.** The corresponding topography maps for 9 clusters. **b.** The grand-averaged waveform partitioning and the waveform representation with selected electrode sites, namely, Pz, Fz, CP5, P6, Cz and CPz. The cluster maps; 6, 8 and 3 are corresponding to P1, P2, and P3 components, and the cluster maps; 5, 2 and 7 are corresponding to N1, N2, and N4 components, respectively. The cluster 9 does not represent any component and the cluster 4 is not a reliable cluster to be considered as a component. Therefore, 6 components are fairly detected by the proposed consensus clustering method.

- *Comparison of the Proposed Method and the Conventional Method in Simulated ERP Data*

To compare the validity of different methods, we calculated the optimal number of clusters from the concatenated dataset using various conventional indices and combined methods. Fig. 5 demonstrates the results of different methods, namely, Elbow, Silhouette, Gap statistic, Dunn, V-fold cross-validation, NbClust R package (i.e. 30 indices within) methods for the simulated ERP. Fig. 5 reveals that the optimal number of clusters is 4 maps by Elbow, 3 maps with Silhouette and NbClust, 11 maps by cross-validation method with k-means from the concatenated data. It is noticeable that Gap statistic and Dunn methods cannot detect the optimal number of clusters from 2-15 clusters. However, observing the results in Fig. 5 shows that it is non-trivial to find a suitable aggregation between the results to identify the optimal number of clusters. To avoid misinterpretation of the results with those different methods, we provided the results of the time-window selection algorithm (i.e. with the proposed method) for the recognized optimal number of clusters obtained from those conventional methods.

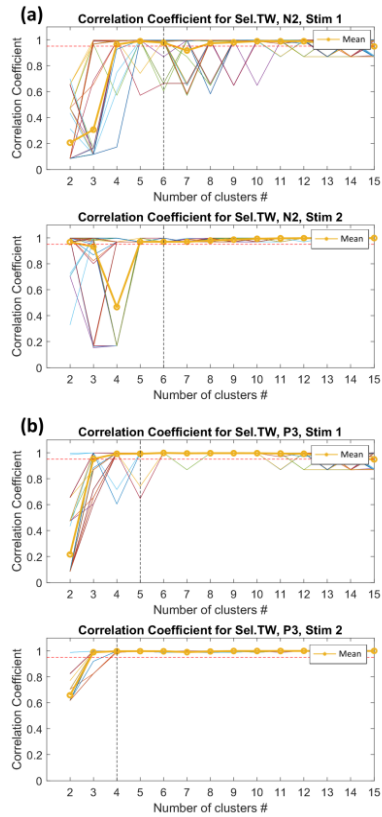

**Fig. 4** Illustration of the optimal number of clusters detection based on the mean of inner-similarity of selected time-windows in 100 runs of the proposed method for N2 and P3 in simulated ERP data. **a.** Detected optimal number of cluster maps for N2 in two stimuli. **b.** Detected optimal number of maps in P3 for two stimuli. The optimal number of clusters is 6 maps for N2 and stimulus 1 and 2, where they are 5 and 4 respectively in P3. The overall optimal cluster maps are detected utilizing the threshold of 0.95 on the mean of recorded inner-similarity (i.e. indicated via orange line) from 100 runs of the proposed method. Together 5 cluster maps is the reasonable option for further analysis. Stim1=stimulus 1, Stim2=stimulus 2.

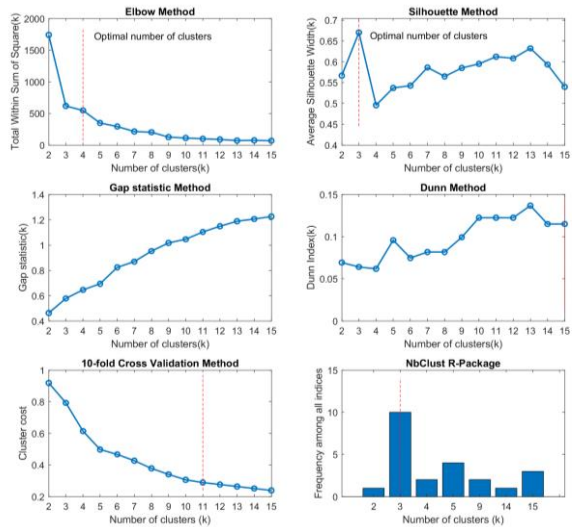

**Fig. 5** Illustration of the optimal number of clusters selection by different popular conventional methods for restructured simulated ERP dataset. A simple observation indicates that the number of the optimal maps are 6 maps in Elbow, 3 maps with Silhouette, 4 maps via Gap statistic, 3 maps in Dunn index, 6 maps in 10-fold cross-validation using k-means, and finally 3 maps for NbClust combined method. There is no aggregation among the different methods for selecting the optimal number of clusters.

- *Cluster Analysis in Simulated ERP*

The clustering results of the proposed method on the simulated ERP data are shown in Fig. 6 and Fig. 7. Fig. 6a,b represent the clustering results in simulated ERP data for two cues (i.e., stimulus1 and stimulus 2) and N2 component. Similarly, Fig. 7a,b reveal the results of clustering for two cues and P3 component. The spatial correlation of time samples and corresponding topographic maps for the selected cluster maps, and the waveforms partitioning for N2 and P3 components are shown in Fig. 6 and Fig. 7 in more details, respectively. Therefore, Fig. 6 illustrates that the N2 component can be isolated by cluster map 3 for both tasks (cue 1 and cue 2) from 217 to 250 ms and 208 to 259 ms, respectively. Similarly, in Fig. 7 the P3 can be isolated via cluster map 3 from 282 to 334 ms and 273 to 343 ms post-stimulus for cue 1 and cue 2, respectively. Interestingly, the high correlation coefficient (i.e. close to 1) within time samples' topographies in predicted time-window for N2 and P3 in both cues can prove our hypothesis about the properties of selected cluster map for a time-window. The cluster maps properties (i.e. start, end, and duration) will be used in further analysis.

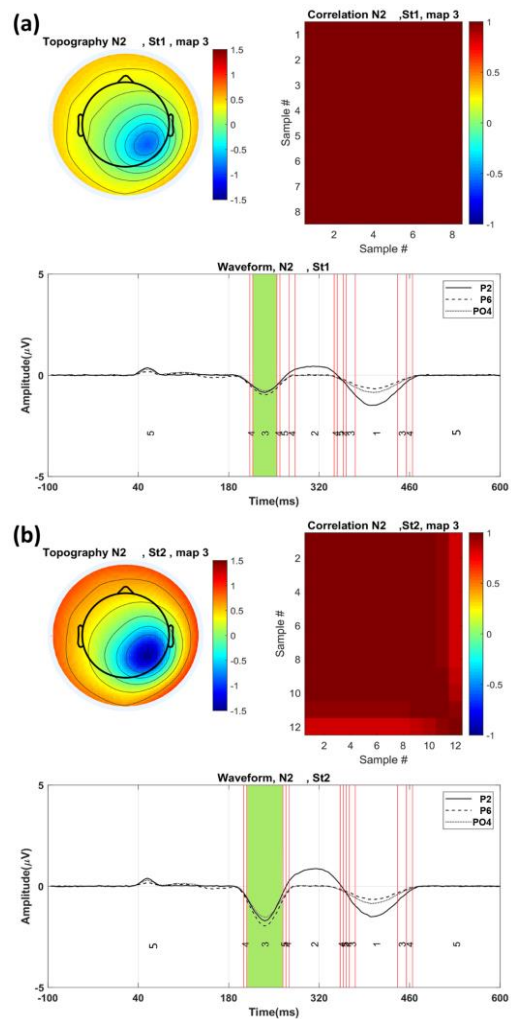

**Fig. 6** The clustering results restructured simulated ERP data (i.e., grand averaged) for N2 component. N2 Component is isolated by microstate map 3 in both stimuli (i.e. stimulus 1 and stimulus 2) from restructured data in simulated ERP data. **a.** Waveform segmentation selected time-window (i.e., duration of 33ms) topography map and correlation of time samples N2 on St1. **b.** Waveform segmentation, selected time-window (i.e., duration of 52ms), topographic map and correlation of time samples for N2 in St2. St1=Stimulus 1 and St2= Stimulus 2.

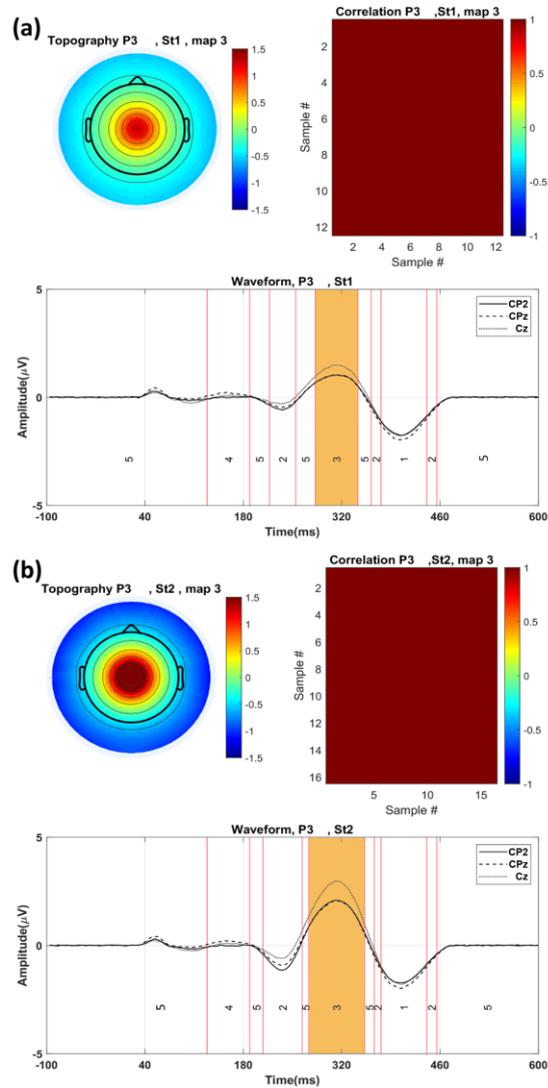

**Fig. 7** The clustering results on restructured simulated ERP data (i.e., grand averaged) for P3 component. P3 component is isolated by the microstate map 3 in both stimuli (i.e. stimulus 1 and stimulus 2) from restructured simulated ERP data. **a.** Waveform segmentation, selected time-window (i.e., duration of 52ms), topography map and correlation of time samples for P3 in St1. **b.** Waveform segmentation, selected time-window (i.e., duration of 71ms), topographic map and correlation of time samples for P3 in St2.

- *Time-window Selection Results in Simulated ERP Data*

The selecting time-windows results of different methods on simulated ERP data are reported in Table 2. Briefly, we demonstrate the predicted time-windows for the optimal number of clusters obtained by various methods for simulated ERP data. Technically, the measurement intervals in the simulated ERP data for two components, namely, N2 and P3 were determined by simulation software. However, the determined measurement windows for N2 and P3 (i.e. with simulated software) and both conditions were roughly from 183 to 278 ms and from

**Table 2** Selected time-windows for various methods obtained in the optimal number of clusters. it represents the selected time-windows with proposed time -window selection method for namely Elbow, Cross-validation, Silhouette, Dunn, NbClust, Gap statistic and X-means for St1 and St2. St1=stimulus 1, St2=Stimulus 2.

| Method | Time-window selection algorithm results for N2 and P3 | | | |
|---|---|---|---|---|
| | TWs for N2 | | TWs for P3 | |
| | St1 | St2 | St1 | St2 |
| NbClust Silhouette | 212-254 | 217-250 | 259-278 | 273-348 |
| Elbow X-means | 217-254 | 208-259 | 282-334 | 273-343 |
| Proposed method | 217-250 | 208-259 | 282-334 | 273-343 |
| Cross-validation | 212-254 | 217-250 | 287-329 | 287-334 |
| Microstates method | 128-194 | 198-259 | 264-348 | 264-357 |

231 to 350 ms post-stimulus, respectively. In this sense, our proposed method carried out the best time-window for N2, 217 to 250 ms in cue 1 and 208 to 259 ms for cue 2. Those time-windows were 212 to 254 ms and 217 to 250 ms for Silhouette and NbClust methods, 217 to 254ms and 208 to 259 ms for Elbow and *X*-means, 212 to 254 ms and 217 to 250 ms for 10-fold cross-validation and, 128 to 194 ms and 198 to 259 ms for Microstates analysis with cross-validation (Table 2). Similarly, in P3, the predicted time-window for cue1 and cue 2 were 282 to 334 ms and 273 to 343 ms with the proposed method, 259 to 278ms and 273 to 348 ms with NbClust and Silhouette methods, 282 to 334 ms and 273 to 343 ms by Elbow and X-means methods, 287 to 329 ms and 287 to 334 ms Cross-validation method and , 264 to 348 ms and 264 to 357 ms via the ERP microstates analysis method. As a result, the predicted time-window with the proposed approach for both cues indicate highly overlapped with the defined components and inner-similarity (higher than 0.95) however, not exactly the same. This caused by the exist overlap within the components which is normal in clustering ERPs.

- *Performance Results in Simulated Data*

We measured the mean amplitude of grand-means data in the selected time-windows (i.e. in Table 2) and defined sensor sites. As we described before, we considered two within-subject factors, *Task* (Stimulus 1 and Stimulus 2) and *Electrode* sites (P2, P6, PO4 in N2 and CP2, CPz, Cz in P3) and interaction between *Task* and *Electrode* for two components of interest in simulated ERP data. Table 3 represents the results of statistical power analysis for simulated ERP data. The results reveal that the main effect of *Task*, *Electrode*, and interaction between *Task* and *Electrode* are significant (p < 0.0001) for both N2 and P3 by all the methods, namely, Silhouette, NbClust, X-means, Elbow, Proposed method, Cross-validation, and Microstates analysis (i.e. for more details see Table 3).

**Table 3** Summarization of selected optimal number of clusters and statistical power analysis results for simulated ERP data in N2 and P3 components. This table demonstrates statistical analysis for various methods in selected time window utilizing time-window selection algorithm (i.e., except microstate analysis methods) in this study (see the text). All the methods could achieve significant difference for different factors in components of interest.

| | | | The optimal number of clusters and statistical analysis results | | | | | |
|---|---|---|---|---|---|---|---|---|
| | | | N2 in P2, P6, PO4 | | | P3 in CP2, CPz, Cz | | |
| Method | Criterion | OPNC | Task | Electrode | Task x Electrode | Task | Electrode | Task x Electrode |
| Silhouette | Objects Similarity | 3 | p<0.0001 F(1,19)=271.38 | p<0.0001 F(2,38)=196.95 | p<0.0001 F(2,38)=93.94 | p<0.0001 F(1,19)=270.84 | p<0.0001 F(2,38)=258.10 | p<0.0001 F(2,38)=263.58 |
| NbClust | 30 indices | | | | | | | |
| X-means | BIC-value | 4 | p<0.0001 F(1,19)=271.35 | p<0.0001 F(2,38)=221.60 | p<0.0001 F(2,38)=82.90 | p<0.0001 F(1,19)=266.20 | p<0.0001 F(2,38)=284.74 | p<0.0001 F(2,38)=211.22 |
| Elbow | Explained variance | | | | | | | |
| Proposed method | Inner-similarity | 5 | p<0.0001 F(1,19)=273.64 | p<0.0001 F(2,38)=220.41 | p<0.0001 F(2,38)=72.53 | p<0.0001 F(1,19)=266.20 | p<0.0001 F(2,38)=284.74 | p<0.0001 F(2,38)=211.22 |
| Cross-validation | Stability | 11 | p<0.0001 F(1,19)=271.38 | p<0.0001 F(2,38)=196.95 | p<0.0001 F(2,38)=93.94 | p<0.0001 F(1,19)=268.75 | p<0.0001 F(2,38)=282.79 | p<0.0001 F(2,38)=236.94 |
| Microstates by Cross-validation | Stability | 11 | p<0.0001 F(1,19)=267.21 | p<0.0001 F(2,38)=199.19 | p<0.0001 F(2,38)=141.56 | p<0.0001 F(1,19)=267.97 | p<0.0001 F(2,38)=278.47 | p<0.0001 F(2,38)=220.15 |

Furthermore, to illustrate the stability and reliability of the proposed method, we conducted 100 iterative calculations of statistical analysis to illustrate the stability of the proposed approach in simulated ERP data. The calculated mean *p*-value for analyzing the interested factors, namely, *Task* and *Electrode* and interaction between them *Task* and *Electrode* for N2 and P3 in simulated ERP are illustrated in Table 4. Table 4 shows suitable stability in the statistical power analysis in calculating the mean *p*-value and the standard deviation for the obtained *p*-values. This reveals that by repeating many times the proposed method the statistical power analysis results are fairly stable and reliable in practice.

**Table 4** Illustration of mean *p*-value calculation in 100 iterative calculations for statistical analysis and STD value across all *p*_value results in N2 and P3 components. STD= Standard error deviation across all iterations.

| | N2 | | |
|---|---|---|---|
| **Criteria** | **Task** | **Electrode** | **Task x Electrode** |
| Mean p-value | 9.59E-13 | 6.23E-21 | 4.11E-14 |
| STD | 6.69E-14 | 1.41E-21 | 1.84E-13 |
| | P3 | | |
| **Criteria** | **Task** | **Electrode** | **Task x Electrode** |
| Mean p-value | 1.22E-12 | 1.56E-23 | 1.84E-19 |
| STD | 6.20E-14 | 1.08E-23 | 1.41E-18 |

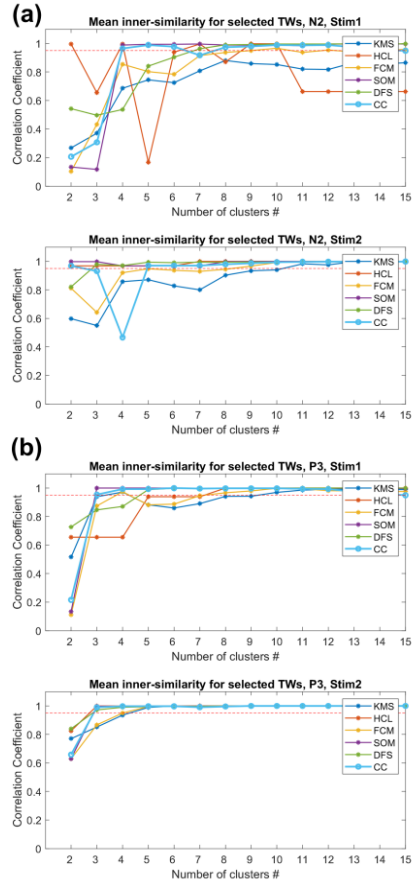

**Fig. 8** Various clustering methods performance regarding inner-similarity of selected time-window for clustering options from 2 to 15 maps in restructured simulated ERP dataset. **a.** Mean inner-similarity of selected time-windows for N2 component isolation from multiple clusterings. **b.** Mean inner-similarity of selected time-windows for P3 component isolation from multiple clusterings. The behavior of different clustering methods in P3 is more correlated compare with clustering for N2. Consensus clustering (blue bold line) has better performance in terms of stability of inner-similarity for selected time-window. KMS=k-means, HCL= Hierarchical clustering, FCM=Fuzzy C-means, SOM=self-organizing map, DFS=Diffusion map spectral clustering, CC=consensus clustering.

- *Contribution of Individual Clustering Method in Simulated ERP Data*

**Table 5** The mutual contribution of clustering methods in Simulated ERP data. Consensus=Consensus clustering, FCM=Fuzzy-C-means, SOM=Self-organizing map, DFS=Diffusion map spectral clustering.

| Clustering method | Consensus | K-means | Hierarchical | FCM | SOM | DFS |
|---|---|---|---|---|---|---|
| **Consensus** | - | 0.976 | 0.863 | 0.866 | 0.958 | 1.000 |
| **K-means** | - | - | 0.840 | 0.844 | 0.943 | 0.976 |
| **Hierarchical** | - | - | - | 0.995 | 0.827 | 0.863 |
| **FCM** | - | - | - | - | 0.832 | 0.866 |
| **SOM** | - | - | - | - | - | 0.958 |
| **DFS** | - | - | - | - | - | - |

For better understanding, the role of each studied clustering method in the proposed consensus clustering, the mean average of inner-similarities via various clustering methods and the number of clusters options are demonstrated in Fig. 8. The comparison between the behavior of mean inner-similarities with different clustering methods in clustering options reveals the role of each clustering method in consensus clustering (i.e, indicated with bold light blue color in the figure). It is observable that the mean inner-similarity of each individual clustering method and consensus clustering followed the similar behavior. However, in P2 component, the clustering method demonstrates similar behavior as shown in Fig. 8a. It is worthwhile to say that this technical problem (i.e. the inner-similarity heterogeneity between clustering methods) has not decreased the performance of the consensus clustering results. Therefore, the consensus clustering performance in this criterion was not less than individual clustering methods. Furthermore, in P3 component, the behavior of different clustering methods was similar and more homogeneous. We demonstrated the contribution of clustering methods Fig.8b for P3 component.

In order to study the mutual contribution of clustering methods to the ensemble clustering, we calculated mutual information of every two clustering methods to exhibit the similarity between the obtained labeling results. *Rand distance* (i.e. a suitable similarity measurement with different labelings for the clusterings) (Strehl and Ghosh 2003) has been used for this aim. Rand index can be explained by the following equation:

$$\mathcal{R}(L, L') = \frac{N_{11} + N_{00}}{n(n-1)/2} \tag{16}$$

where $N_{00}$ is the number of object pairs that are clustered in the different clusters in $L$ and $L'$. While $N_{11}$ is the number of object pairs that are in the same clusters in $L$ and $L'$. The contribution of each of these clustering methods to simulated ERP data is shown in Table 5. Table 5 reveals that the maximum contribution of the clustering methods was 1.000 obtained with Consensus and Diff-Spectral methods, where the minimum was 0.863 with consensus clustering and hierarchical clustering methods. The mean contribution of all clustering methods was 0.932. The contribution test shows a reasonable contribution carried out by all the studied clustering methods in the consensus clustering.

### 3.2 Results For real ERP data

- *Inner Similarity Analysis and the Optimal Number of Clusters Selection for Real ERP Data*

It is observable in Fig. 9, the inner-similarity of the selected time-windows is increasing after few unstable states (i.e., mostly before 4 or 5 cluster maps), however, this increase is smooth after 5 clusters in order to set out threshold of inner-similarity (i.e. 0.95). In spite of Fig. 9a and HC which the proposed algorithm selects 10 cluster maps for both tasks (i.e., because of high threshold limit), according to Fig. 9a, 5 cluster maps can be chosen as the optimal number of clusters for N300 and RS group. Likewise, in Fig. 9b the optimal numbers of clusters for prospective positivity in RS and PM and Ongoing tasks are 5 and 4 maps, respectively, which they are 6 and 7 cluster maps in HC group and PM and Ongoing tasks, correspondingly. Therefore, 6 cluster maps can be a reasonable choice for prospective positivity component analysis. Together, the proposed method selects 6 cluster maps as the optimal number of clusters to satisfy the level of threshold and stability. It is also worthwhile to mention, by increasing the number of clusters, the inner-similarity value for selected time-windows might converge to higher values, but not necessarily reliable because of relatively thinner clusters with fewer temporal samples and higher inner-similarity.

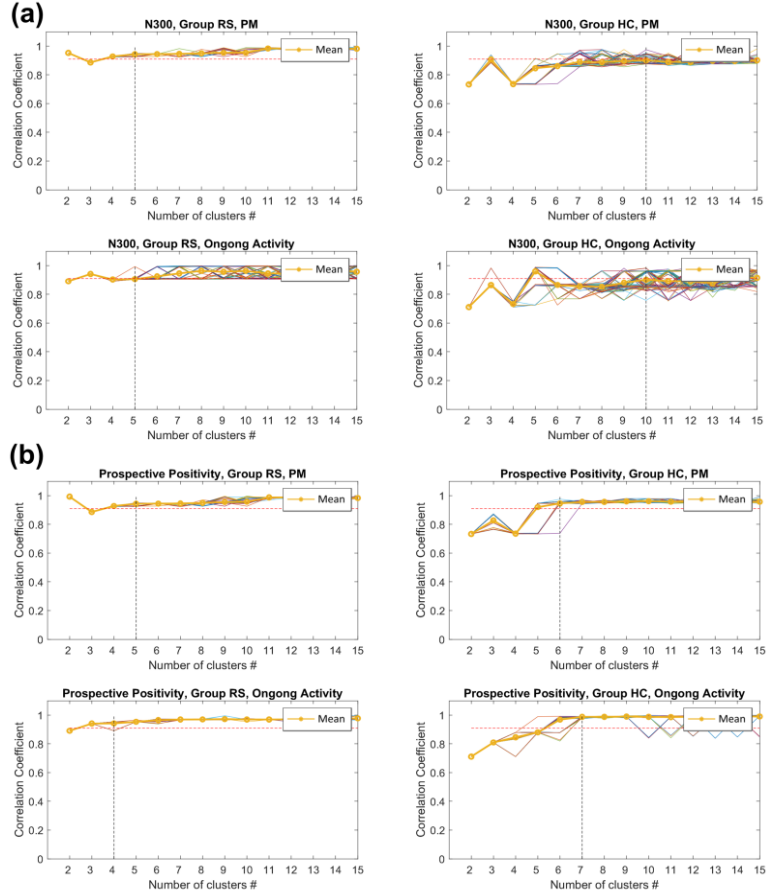

**Fig. 9** The optimal number of clusters detection based on the mean of inner-similarity of selected time-windows in 100 runs of the proposed method for N300 and Prospective positivity components in real data (i.e., Prospective positivity experiment ERP). **a.** The optimal number of clusters for N300. **b.** The optimal number of clusters for prospective positivity. The optimal number of clusters revealed in 5 cluster maps in N300 for all conditions/groups while it was 6 maps for PM task and 7 maps in ongoing activity. The overall optimal cluster maps is detected utilizing the threshold of 0.95 on the mean of recorded inner-similarity (i.e. indicated via orange line) from 100 runs of the proposed method. Together 6 cluster maps is the reasonable option for further analysis.

- *Comparison of the Proposed Method and the Conventional Method for Real ERP Data*

For better comparison, we calculated the optimal number of clusters from concatenated ERP datasets using various conventional methods. Observing Fig. 10, the obtained optimal number of clusters is 6 maps using Elbow, 4 maps via Gap statistic and, 3 maps with Silhouette, Dunn and NbClust package, and 6 maps using Cross-validation for Real ERP concatenated data. Overall, in the results of Fig. 10 seems to be non-trivial to find a suitable aggregation between the results to identify the optimal number of clusters in real ERP data. This happens in the cause of the effect of dataset properties in selecting the optimal number of clusters by the different methods. It is needless to mention that the results of the time-window selection in the recognized optimal number of clusters have been calculated by the proposed time-window selection method.

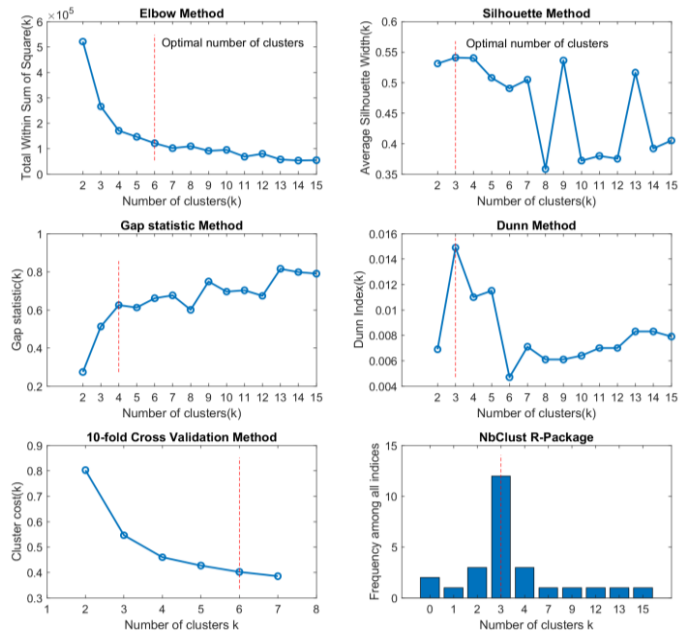

**Fig. 10** Illustration of the optimal number of clusters selection by different popular conventional methods for restructured ERP dataset for PM experiment. A simple observation indicates that the number of the optimal maps were 6 maps in Elbow, 3 maps with Silhouette, 4 maps via Gap statistic, 3 maps in Dunn index, 6 maps in 10-fold cross-validation using k-means, and finally 3 maps for NbClust combined method. No aggregation can be found among the different optimal number of clusters is observed.

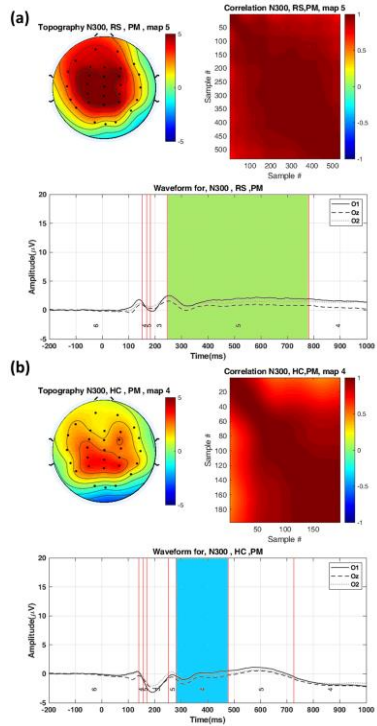

**Fig. 11** Demonstration of clustering results and microstate maps for isolating N300 component from restructured ERP in two groups (i.e., map 5 for RS and map 4 for HC) for PM task and segmentation with consensus clustering. **a.** Topographic map and correlation between time samples in the selected time-window (i.e., duration of 534ms), in RS group and PM task. **b.** Selected time-window (i.e., duration of 197ms), topographic map and correlation of time samples for isolated N300 in HC and PM.

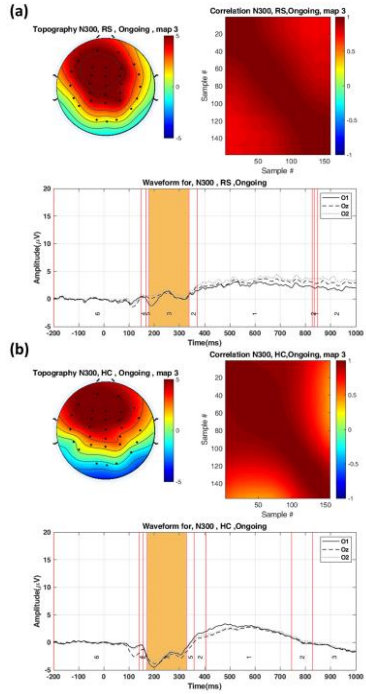

**Fig. 12** Demonstration of clustering results and microstate maps for isolating N300 component from restructured ERP in two groups (i.e., RS and HC) for Ongoing task and segmentation by consensus clustering. **a.** Shows the corresponding ERP waveforms, the selected time-window (i.e., duration of 158ms), the topographic map and correlation for corresponding time samples. **b.** Waveform partitioning, selected time-window (i.e., duration of 156ms), topographic map and correlation for corresponding time samples.

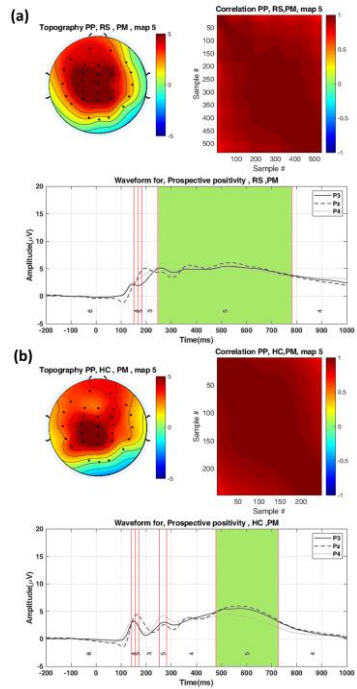

**Fig. 13** Illustration of ERP waveform segmentation and isolating prospective positivity component (i.e., map 5 for both RS and HC groups) in PM task from restructured ERP. **a.** Selected time-window (i.e., duration of 534ms), topographic map and corresponding correlation for time samples in the selected time-window in RS and PM task. **b.** Selected time-window (i.e., duration of 247ms), topographic map and correlation of time samples for isolated prospective positivity in HC and PM task.

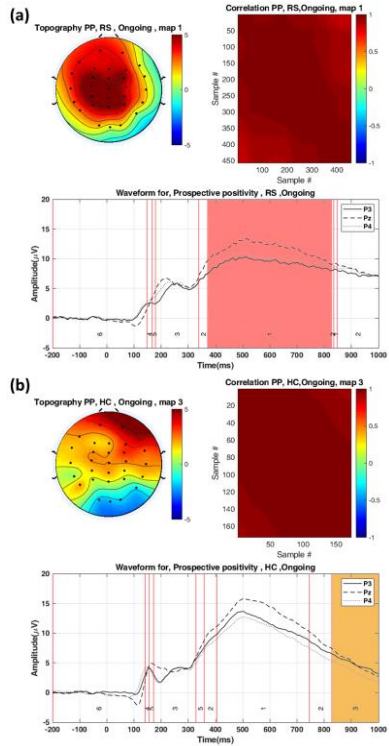

**Fig. 14** Illustration of isolating prospective positivity component from restructured ERP for two groups (i.e., map 1 for RS and map 3 for HC) regarding the ongoing task. **a.** Topographic map and correlation of time samples in the selected time-window (i.e., duration of 456ms) in RS group and ongoing task. **b.** The time-window (i.e., duration of 172ms), the topographic map and correlation of the time samples for isolated Prospective positivity in HC and ongoing task.

- *Cluster Analysis for Real ERP Data*

Fig. 11 illustrates the results for PM task and two groups regarding the isolated N300 component applying the optimal number of clusters (i.e., 6 cluster maps). It is observable in Fig. 11a that the time-window for measuring N300 in RS group extracted by cluster map 5 is isolated from 246 to 779 ms post-stimulus. Likewise, in HC group, N300 is isolated by map 4 from 281 to 477 ms. The color marked area represents the time-window, as well as the topographic map and the spatial correlation between samples for the extracted N300 via indicated cluster maps, are shown for each group in Fig.11. Accordingly, Fig. 12 represents the N300 component for both RS and HC of Ongoing task isolated via cluster map 3. Therefore, the time-window for RS group is from 179 to 336 ms, whereas it is 172 to 327 ms for HC group according to the results in Fig. 12. Fig. 13 gives us information about prospective positivity component isolation for PM task in two groups. The prospective positivity is isolated by cluster map 5 for both RS and HC groups from 246 to 779 ms and 478 to 724 ms for RS and HC, respectively. Similarly, Fig. 14 shows the extracted prospective positivity via map 1 in Fig. 14a and map 3 in Fig. 14b for RS and HC, respectively. Hence, the prospective positivity can be measured in 371 to 826 ms in RS group and 828 to 999 ms in HC group. In this sense, the important finding from these four figures (i.e., Fig. 11 to Fig. 14) is that the quasi-stable topography in the selected cluster map can be obtained based on two criteria (i.e. inner similarity and overlap ratio with defined time-window by the experimenter).

- *Time-window Selection Results for Real ERP Data*

Table 6 includes the measurement time-windows based on the predicted optimal numbers of clusters for the proposed method, prior study, and conventional methods, namely, Elbow, Silhouette, Gap statistic, Dunn, v-fold cross-validation, NbClust, *X*-mean, and Microstates analysis in real ERP data. We have exhibited the selected cluster maps and corresponding time interval in the above subsection (i.e. Cluster analysis for real ERP). For instance, the selected time window for measuring N300, RS group and PM cue were 246 to 779 ms and 179 to 336 ms for Ongoing task in RS group by the proposed method, Elbow and cross-validation methods, whereas it was 199 to 400 ms on both tasks in the prior study. This interval was 158 to 999 ms for PM and 339 to 999 ms for the Ongoing task in Silhouette, Dunn, NbClust methods. Gap statistic, X-means select 262 to 999 ms for PM and 177 to 359 ms for ongoing task. Finally, Microstates analysis used modified K-means (Murray et al. 2008) selected time-window in 271 to 999 ms for PM and 167 to 322 ms for Ongoing task (i.e., for more detail information see Table 6).

**Table 6** Illustration of selected time-window in studied methods in the obtained optimal number of clusters. This table gives the selected time-windows with proposed time-window selection method for namely Elbow, Cross-validation, Silhouette, Dunn, NbClust, Gap statistic and X-means in N300 and Prospective positivity and different groups (i.e., Remitted schizophrenia (RS) and Healthy controlled (HC)), two paradigms Prospective memory (PM) and ongoing activity.

| Method | Time-window selection algorithm results for N300 and Prospective positivity | | | | | | | | | | | |
|---|---|---|---|---|---|---|---|---|---|---|---|---|
| | TWs for N300 | | | | | | TWs for Prospective positivity | | | | | |
| | RS | | HC | | | | RS | | HC | | | |
| | PM | Ongoing | PM | Ongoing | | | PM | Ongoing | PM | Ongoing | | |
| Prior Study | 190-400 | 190-400 | 190-400 | 190-400 | | | 400-1000 | 400-1000 | 400-1000 | 400-1000 | | |
| Proposed method | | | | | | | | | | | | |
| Elbow | 246-779 | 179-336 | 281-477 | 172-327 | | | 246-779 | 371-826 | 478-724 | 821-999 | | |
| Cross-validation | | | | | | | | | | | | |
| Silhouette | | | | | | | | | | | | |
| Dunn | 158-999 | 339-999 | 146-309 | 270-362 | | | 158-999 | 339-999 | 336-960 | 363-875 | | |
| NbClust | | | | | | | | | | | | |
| Gap statistic | 262-999 | 177-359 | 253-990 | 170-376 | | | 262-999 | 360-922 | 253-990 | 377-783 | | |
| X-means | | | | | | | | | | | | |
| Microstates method | 271-999 | 167-322 | 271-999 | 167-322 | | | 271-999 | 328-999 | 271-999 | 328-999 | | |

- *Performance Results for Real ERP Data*

The statistical power analysis for the main effect of two within-subject factors *Task* (PM and Ongoing), *Electrode* sites ($O_1, O_z$ and $O_2$ for N300 and $P_3$, $P_z$ and $P_4$ for prospective positivity) and between-subject factor, *Group* (RS and HC) and also interaction between *Task* and *Group* is reported in this section. Table 7 illustrates the results of statistical analysis from various methods in real ERP. First of all, the main effect of *Task* type in prior study (F(1,227)=6.43, p < 0.05) , the propose method, Elbow and Cross-validation (F(1,38)=9.2133, p < 0.004), NbClust, Dunn and Silhouette methods (F(1,38)=38.00, p < 0.021), Gap statistic and *X*-means methods (F(1,38)=8.88, p < 0.005) and Microstates analysis with modified k-means (F(1,38)=27.11, p < 0.0001) was significant in all the methods in N300 component. The microstates analysis indicates the highest difference compared to all the other studied methods between two tasks waveform. Also, the main effect of *Group* in prior study (F(1,227)=89.24, p < 0.001), propose method, Elbow and Cross-validation (F(1,38)=9.56, p < 0.004), NbClust, Dunn and Silhouette methods(F(1,38)=11.566, p < 0.002), Gap statistic and *X*-means methods (F(1,38)=7.548, p < 0.009) and (Microstates analysis F(1,38)=9.22, p < 0.004) was significant in all above studied methods, suggesting the prior study is detects higher significant difference between two groups, comparatively. Moreover, the main effect of *Electrode sites* in prior study ($F(2,227)=3.30$, $p < 0.05$), propose method, Elbow and Cross-validation($F(2,76)=3.00$, $p < 0.05$), NbClust, Dunn and Silhouette ($F(2,76)=3.30$, $p < 0.04$) methods was significant unlike Gap statistic and X-means methods ($F(2,76)=2.77$, $p < 0.07$) and Microstates analysis ($F(2,76)=2.83$, $p < 0.07$). Finally, the interaction effect between *Task* and *Group* in the propose method, Elbow and Cross-validation ($F(1,38)=8.41$, $p < 0.006$), NbClust, Dunn and Silhouette ($F(1,38)=38.00$, $p < 0.010$), Microstates analysis ($F(1,38)=12.25$, $p < 0.001$) methods was significant opposite to prior study, Gap statistic and *X*-means methods ($F(1, 38)=1.471$, $p < 0.233$).

Table 7 The summarized optimal number of clusters detection using different methods for PM experiment. This table gives an overview of statistical power analysis comparison results in N300 and Prospective positivity components. OPNC= Optimal number of clusters, NS= not significant.

| Method | Criterion | OPNC | Range | The optimal number of clusters and statistical analysis results | | | | | | | |
|---|---|---|---|---|---|---|---|---|---|---|---|
| | | | | N300, O1, Oz, O2 | | | | | Prospective positivity, P3, Pz, P4 | | |
| | | | | Task | Group | Electrode | Task x Group | Task | Group | Electrode | Task x Group |
| Prior Study | Subjective | - | - | $p < 0.05$ $F(1,227)=6.43$ | $p < 0.001$ $F(1,227)=89.24$ | $p < 0.05$ $F(2,227)=3.30$ | NS | $p < 0.001$ $F(1,227)=250.61$ | $p < 0.005$ $F(1,227)=4.62$ | $p < 0.01$ $F(2,227)=5.44$ | $p < 0.01$ $F(1,227)=10.19$ |
| Proposed method | Inner similarity | 6 | 2-15 | | | | | | | | |
| Elbow | Explained variance | 6 | 2-15 | $p<0.004$, $F(1,38)=9.21$ | $p<0.004$ $F(1,38)=9.56$ | $P<0.05$ $F(2,76)=3.00$ | $p<0.006$ $F(1,38)=8.41$ | $p<0.0001$ $F(1,38)=75.40$ | $p<0.0001$ $F(1,38)=34.90$ | $p<0.033$ $F(2,76)=3.54$ | $p=0.003$ $F(1,38)=9.46$ |
| Cross-validation | Stability | 6 | 2-15 | | | | | | | | |
| Silhouette | Objects Similarity | 3 | 2-15 | | | | | | | | |
| Dunn | Distance from mean | 3 | 2-15 | $p<0.021$, $F(1,38)=38.00$ | $p<0.002$ $F(1,38)=11.566$ | $p<0.04$ $F(2,76)=3.30$ | $p<0.010$, $F(1,38)=38.000$ | $p<0.0001$, $F(1,38)=233.89$ | **$p<0.903$ $F(1,38)=0.015$** | $p<0.002$ $F(2,76)=6.81$ | $p<0.018$, $F(1,38)=6.096$ |
| NbClust | 30 indices | 3 | 2-15 | | | | | | | | |
| Gap statistic | within-group dispersion | 4 | 2-15 | $p<0.005$, $F(1,38)=8.88$ | $p<0.009$ $F(1,38)=7.548$ | **$p<0.07$ $F(2,76)=2.77$** | **$p<0.233$, $F(1,38)=1.471$** | $p<0.0001$, $F(1,38)=286.34$ | **$p<0.902$ $F(1,38)=0.015$** | $p<0.006$ $F(2,76)=5.49$ | $p<0.002$, $F(1,38)=11.292$ |
| X-means | BIC-value | 4 | 2-15 | | | | | | | | |
| Microstate by Cross-validation | Stability | 6 | 2-15 | $p<0.0001$ $F(1,38)=27.11$ | $p<0.004$ $F(1,38)=9.22$ | **$p<0.07$ $F(2,76)=2.83$** | $p<0.001$ $F(1,38)=12.25$ | $p<0.0001$ $F(1,38)=13.05$ | **$p<0.18$ $F(1,38)=1.88$** | $p<0.0007$ $F(2,76)=7.92$ | $p<0.01$ $F(1,38)=6.75$ |

According to statistical power analysis for *prospective positivity* component, the main effect of *Task* type in prior study ($F(1,227)=250.61$, $p < 0.001$), the propose method, Elbow and Cross-validation($F(1,38)=75.40$, $p < 0.0001$), NbClust, Dunn and Silhouette methods ($F(1,38)=233.89$, $p < 0.0001$), Gap statistic and *X*-means methods($F(1,38)=286.34$, $p < 0.0001$) and Microstates analysis ($F(1,38)=13.05$, $p < 0.0001$) was significant in all the methods. Also , the main effect of *Group* prior study ($F(1,227)=4.62$, $p < 0.005$) and propose method, Elbow and Cross-validation ($F(1,38)=34.90$, $p < 0.0001$) was significant which it was not significant in NbClust, Dunn and Silhouette ($F(1,38)=0.015$, $p < 0.903$), Gap statistic and *X*-means methods ($F(1,38)=0.015$, $p < 0.902$) and Microstates analysis ($F(1,38)=1.88$, $p < 0.18$). Moreover, the main effect of *Electrode sites* in prior study ($F(2,227)=5.44$, $p < 0.01$), the propose method, Elbow and Cross-validation ($F(2,76)=3.54$, $p < 0.033$), NbClust,

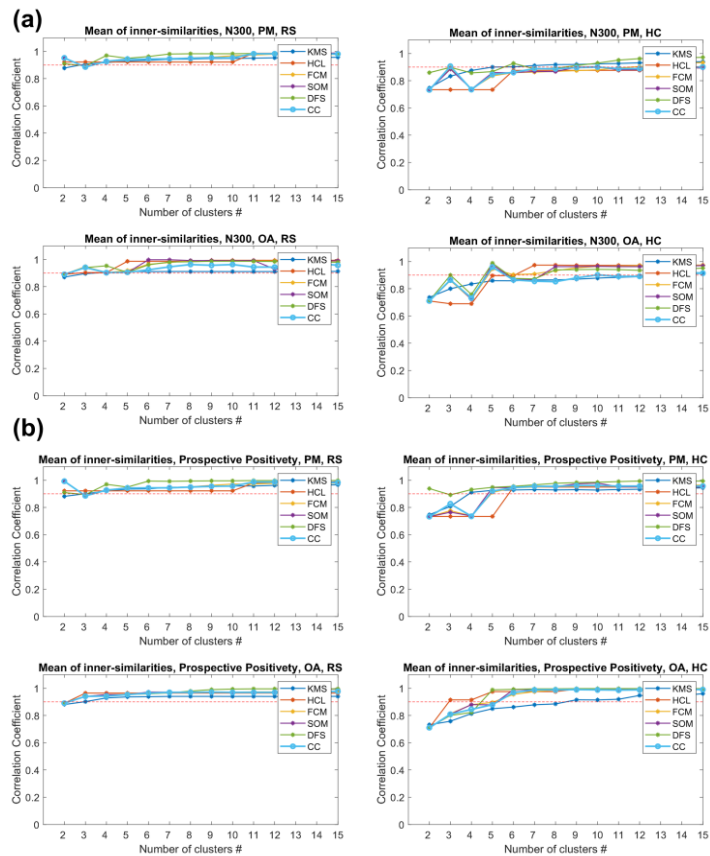

**Fig. 15** Different clustering method performance and contribution for consensus clustering. **a.** Mean inner-similarity for selected time-windows for N300 component. **b.** Mean inner-similarity for selected time-windows for P3 component. The multiple-clustering methods mean inner-similarity for selected time-windows reveals consistency and contribution of each clustering method in final consensus clustering results (blue bold line). KMS=k-means, HCL= Hierarchical clustering, FCM=Fuzzy C-means, SOM=self-organizing map, DFS=Diffusion map spectral clustering, CC=consensus clustering.

**Table 8** Result of 100 iterative calculations for statistical power analysis for N300 and Prospective positivity STD value across all *p*-value results. STD= Standard error deviation across all iterations.

| | N300 | | | |
|---|---|---|---|---|
| **Criteria** | **Task** | **Group** | **Electrode** | **Task x Group** |
| Mean p-value | 3.36E-02 | 1.88E-02 | 4.94E-02 | 2.64E-02 |
| STD | 2.58E-01 | 4.73E-02 | 1.98E-02 | 4.19E-02 |
| | **Prospective positivity** | | | |
| **Criteria** | **Task** | **Group** | **Electrode** | **Task x Group** |
| Mean p-value | 3.88E-07 | 2.70E-03 | 2.65E-02 | 5.14E-02 |
| STD | 0.00E+00 | 2.39E-02 | 1.28E-02 | 1.36E-01 |

Dunn and Silhouette methods ($F(2,76)=6.81$, $p < 0.002$), (Gap statistic and *X*-means methods $F(2,76)=5.49$, $p < 0.006$) and Microstates analysis ($F(2,76)=7.92$, $p < 0.0007$) was significant in all methods. Importantly, the interaction effect between *Task* and *Group* in prior study ($F(1,227)=10.19$, $p < 0.01$), the propose method, Elbow and Cross-validation ($F(1,38)=9.46$, $p < 0.003$), NbClust, Dunn and Silhouette methods ($F(1,38)=6.096$, $p < 0.018$) Gap statistic and *X*-means methods ($F(1,38)=11.292$, $p < 0.002$) and using Microstates analysis ($F(1,38)=6.75$, $p < 0.01$) was significant in all methods. Aim to test the reliability of the statistical analyzing results, we also conducted 100 iterative calculations of statistical analysis to illustrate the stability of the proposed approach in real ERP data. Hence, the calculated mean *p*-value for analyzing the interested factors, namely, *Group*, *Task*, *Electrode*, and interaction between *Task* and *Group* in N300 and prospective positivity data are presented in Table 8.

- *Contribution of Individual Clustering Method for Real ERP Data*

Fig. 15 exhibits similar behavior of the mean of inner-similarities of the selected time-windows after many times running various clustering methods in all conditions/groups for N300 and prospective positivity components. In Fig. 15a, we observe that the mean inner-similarity of each individual clustering method and consensus clustering (i.e., indicated with bold light blue line) followed the similar behavior in N300 component for different groups/conditions. As we expected, the performance of studied clustering methods is close to the proposed consensus clustering method. This reveals that the heterogeneity between clustering methods synergistically supports the performance of consensus clustering results. Therefore, the consensus clustering performance in this criterion was not less than individual clustering methods. Likewise, Fig. 15b illustrates the similar behavior of studied clustering methods with the proposed consensus clustering method involving different groups/conditions. This can be attributed to the role of each individual clustering method in the proposed consensus clustering approach for real ERP data. Moreover, the mutual contribution between every two studied clustering methods is illustrated in Table 9. As can be seen from the table, the proposed method offered a fairly high contribution to real ERP data among other studied clustering methods. The consensus clustering shared maximum mutual information with hierarchical clustering (0.770) and minimum mutual information with diffusion map spectral clustering (0.748) with an average of 0.763.

**Table 9** The mutual contribution of clustering methods in real-related ERP data. Consensus=Consensus clustering, FCM=Fuzzy-C-means, SOM=Self-organizing map, DFS=Diffusion map spectral clustering.

| Clustering method | Consensus | K-means | Hierarchical | FCM | SOM | DFS |
|---|---|---|---|---|---|---|
| **Consensus** | - | 0.762 | 0.770 | 0.769 | 0.766 | 0.748 |
| **K-means** | - | - | 0.936 | 0.934 | 0.913 | 0.875 |
| **Hierarchical** | - | - | - | 0.998 | 0.935 | 0.886 |
| **FCM** | - | - | - | - | 0.934 | 0.885 |
| **SOM** | - | - | - | - | - | 0.868 |
| **DFS** | - | - | - | - | - | - |

## 4. Discussion

This study aims to attain a better understanding of using consensus clustering to determine the optimal number of clusters for the objective prediction of the time window for an ERP of interest. The proposed method then assists to objectively measure the mean over amplitudes of the time window to represent the amplitude of an ERP for cognitive neuroscience (Luck 2014). It demonstrated that ERP analysis results can be affected by clustering the ERP data into a different number of clusters. The results revealed that by utilizing a proper number of clusters, the inner-similarity of topographies of the multiple time samples of the selected time window for extracting the ERP components of interest achieved in the acceptable threshold (i.e., higher than 0.95 in this study). More precisely, the inner-similarity of topographies of the multiple time samples of the found time-window was used as an important criterion for measuring the quality of isolated components and finding the optimal number of cluster maps for spatio-temporal ERP. To evaluate the performance of the proposed method, we utilized the new approach on the simulated ERP data generated by Berg's Dipo software (Berg 2006). To make better sense of interpretation of obtained results compared to previous ERP findings (Chen et al. 2015), we also utilized real ERP data for evaluating and comparing the results of the new approach with conventional methods. As expected, the results of the proposed method on simulated and real ERP data revealed the reliability of obtaining the optimal number of clusters for spatio-temporal ERP.

Consequently, one should be aware that only the magnitudes of inner-similarity of candidate time-windows cannot be sufficient criterion for selecting the best time-window during the isolation of the component of interest. In other words, one may find a microstate map (cluster map) with the highest satisfaction of the inner-similarity, however, it may not be able to isolate the component of interest. This issue can be more confusing when the high resolution of cluster maps is employed for clustering (i.e., clustering with a higher number of clusters). As an example, to clarify this issue, in Fig. 4a, b the inner-similarity of the selected time window is extremely high after a specific number of clusters, e.g., 7 clusters and etc. This can mislead the researchers to select a higher number of clusters. As a result, many thin clusters might be generated by clustering, thus, the selected time-windows would not be able to properly represent the components of interest. Therefore, we considered two criteria for determining the best time-window: Firstly, we selected two high-ranked clusters according to the experimentally defined interval (i.e., by the experimenters) with the satisfactory threshold of the interval duration (i.e., minimum acceptable duration). This entails that the defined interval included only a few topographies which can represent the component of interest. Therefore, we set the threshold of 30 ms for selecting candidate clusters in simulated and real ERP data because, a reliable component has a reasonable duration in ERP in practice (Kappenman and Luck 2012). This can help to avoid noisy clusters or non-component clusters. Secondly, we calculated the overlap ratio with the defined interval by the experimenter for selecting outperformed time-windows.

The other critical issue in the consensus clustering is to perform a proper combination of clustering methods for ensembling task. Indeed, the main rationale of using consensus clustering technique is to provide the facility of combining the specific advantages of multiple clustering algorithms and reliability without decreasing the performance (Basel Abu-Jamous 2015; Liu et al. 2017a; Nguyen and Caruana 2007; Strehl and Ghosh 2003; Topchy et al. 2004; Vega-Pons and Ruiz-Shulcloper 2011). Therefore, it is necessary to consider the performance of the single clustering algorithms before employing it in an ensembling process. The weak clustering results might lead to unsatisfactory results and misinterpretations. In this study, we successfully applied the standard version of K-means, fuzzy C-means (FCM) (Dinov and Leech 2017), hierarchical clustering, modified K-means (Koenig et al. 2014; Michel and Koenig 2018; Murray et al. 2008; von Wegner et al. 2018) which indicated that these clustering methods can be used for clustering spatio-temporal ERP. As a consequence, the self-organizing map (SOM) (Abu-Jamous et al. 2014) and diffusion map spectral clustering (already used in fMRI data processing) (Sipola et al. 2013b) is also useful for processing EEG/ERP data.

For the sake of demonstrating the contribution of each individual clustering method, we illustrate the mean inner-similarity of the selected time-windows for each clustering method in Fig. 8 and Fig. 15 for simulated and real ERP data, respectively. It is observable that, all clustering methods achieve similar performance regarding the inner-similarity criterion. Moreover, we have also tested the statistical power analysis in 100 runs. Table 4 represents the means *p*-value and standard error by *p*-values for 100 runs in simulated data for N2 and P3 components. Equally, the results from the mean *p*-value and standard error by *p*-values in the real experiment, N300, and prospective positivity components were shown in Table 8. This indicates that the results of statistical power analysis for two ERP data are fairly reliable due to the application of the proposed method many times. Interestingly, the results in Table 5 and Table 9 reveal that various clustering methods mutually provide the ensemble of the clusterings results. Therefore, our finding suggests an efficient method to investigate the ensemble of standard clustering methods to a suitable combination for spatio-temporal ERP analysis.

Technically the important parameters for a brain response such as; the starting time point, the ending time point, and duration of a response can be explained with the ERP microstate map concept (Lehmann et al. 1987). The microstates represent dynamic changes in topographies of the averaged ERP waveform, continuously in a spatial resolution (Lehmann et al. 2009) and the single-trial EEG waveform for extracting the interesting components (Khanna et al. 2015; Tzovara et al. 2012). Reliability and success of the proposed method can also be discussed by the fact that the microstates analyzing methods (Brunet et al. 2011; Koenig et al. 2014; Michel and Koenig 2017; Michel and Koenig 2018; Murray et al. 2008; Pascual-Marqui 2002) are complementary to our approach in terms of isolating the components of interest in both ERPs. We consider the interdependent behavior of these methods by two aspects: Firstly, the predicted time-windows with proposed time-window selection algorithm for both methods (i.e., the proposed approach and microstates analysis method) yielded the similar outcome in both real and simulated ERP data as shown in Table 2 and Table 6. The microstates analysis is compatible with the proposed method in terms of predicted time-windows for N2 and P3 components in simulated ERP data, for example, the isolated P3 component is detected with 284 to 334 ms by the proposed method that is 264 to 348 ms with microstates analysis (Table 2). Table 6 reveals that the N300 component has been predicted by 246 to 779 ms post-stimulus by the proposed approach where it was 271 to 999 ms by microstates analysis method in RS group and PM task. The selected time-windows were similar in RS group and Ongoing task for the proposed method and microstates analysis i.e., 179 to 336 ms and 167 to 322 ms, respectively. Likewise, we found the similar time-windows for proposed method and microstates analysis methods in HC group and two tasks (i.e. see Table 6). Furthermore, in predicting prospective positivity, the time-windows for selected microstate maps are similar to each other for RS in both PM and ongoing tasks for the proposed method and microstates analysis. However, those intervals are not complementary for HC in both PM and ongoing tasks. Practically, this issue is normal when the different number of clusters are used in various methods to predict the time-window in the same dataset. The essential difference between two methods is that the microstates analysis assigns the microstates to predetermine constant maps which cannot be always a reliable choice when dealing with different datasets. In contrast, the proposed approach uses a data-driven approach to find the proper number of clusters regarding the characteristics of the datasets. Since there are always overlapped components included in the real data, clustering for isolating several components is not always reasonable (Keil et al. 2014). Therefore, further work is required to check that the combination of both methods for analyzing a few components could be beneficial and comprehensive.

Secondly, Table 3 and Table 7 exhibit that the proposed method detected the significant main effect in *Task*, *Electrode* and in the interaction between *Task* and *Electrode* factors in simulated ERP data respect N2 and P3 components detection. Similarly, other studied methods in Table 3 achieved a suitable performance for those factors. On the other hand, the proposed method also detected the significant main effect of *Task* and *Electrode* for both components of interest (i.e. N300 and prospective positivity) in real ERP (Table 7). Interestingly, we also found the significant main effect of *Group* and interaction between *Task* and *Group* in both components of interest. Comparison with other methods (i.e. especially microstates analysis), the proposed method outperforms conventional methods in detecting the significant difference in studied factors. It is worthwhile mentioning that in some of the factors the purposed method *p*-value does not achieve the best choice among the other methods. However, unlike our method, the other methods have failed to detect the difference for *Electrode*, the interaction between *Task* and *Group* in N300 and for *Group* in prospective positivity.

In summary, our results in this study extend previous research findings of cluster analysis on EEG/ERP in terms of using consensus clustering technique and the optimal number of clusters detection. Besides, the proposed approach could be welcomed to be used in ERP due to two main reasons: Firstly, analyzing spatio-temporal ERP by focusing on few reasonable components of interest instead of fitting microstates to predetermine microstate maps which have been discussed by Michel et al., (Michel and Koenig 2018). Secondly, despite conventional methods, which have used a single clustering method (Charrad et al. 2014; Kaufman and Rousseeuw 1987; Lleti et al. 2004), the proposed method utilizes a powerful cluster ensembling to represent neurophysiologically interpretable components of interest in data. Finally, this study demonstrates a soft combination of neuroscience and machine learning in terms of designing a data-driven based solution for ERP analyzing. As a consequence, in line with numerous significant studies which have investigated segmentation of EEG/ERP with different clustering algorithms (Brunet et al. 2011; Dinov and Leech 2017; Khanna et al. 2015; Koenig et al. 2014; Lehmann 1989; Lehmann 1990; Michel and Koenig 2017; Michel and Koenig 2018; Murray et al. 2008; Pascualmarqui et al. 1995; Ruggeri et al. 2019; von Wegner et al. 2018), the proposed method is appropriate for single-trial EEG because the runs can be considered as the number of trials for each condition/group or in multi-dimensional analysis. Therefore, selecting the number of reliable microstate maps for single-trial EEG can be investigated with the proposed approach. This can be considered in more dimensions and detail analysis to investigate the complexity of processing EEG/ERP. In order to provide the access to the new methodology, a toolbox has been developed under MATLAB platform named OptNC_ERP toolbox (Mahini and Cong 2019) available on simulated ERP data (i.e. unpublished data), which can be used beside EEGLAB (Delorme and Makeig 2004).

## 5. Conclusions and Outlook

This study exhibited the successfully utilizing consensus clustering on spatio-temporal event-related potentials (ERPs) for predicting the most suitable time-window for the component of interest. Our results revealed that, studied various standard clustering methods, namely, K-means, hierarchical clustering, fuzzy C-means (FCM), self-organizing map (SOM) and diffusion map spectral clustering can be successfully combined in a synergistic cluster ensemble manner for extracting the interesting microstate map which can isolate the component of interest. We investigated the dynamic properties of ERP microstate maps (i.e. special correlation of microstates sequences) and temporal properties (i.e. start, end, and duration) to predict measurement time-window for an ERP. The new procedure thus seems to improve the statistical power analysis results by detecting the most suitable time-windows in conditions/groups. Indeed, the new approach discovers a reliable and fairly stable brain response based on real topography state analyzing. We would, therefore, finish by declaring that the results might furthermore offer a starting to the deep research on electroencephalogram (EEG) as a rich source of information and true neuroimaging method (Biasiucci et al. 2019; Michel and Brunet 2019; Michel and Murray 2012). We, therefore, believe that the EEG neuroimaging method can be studied via consensus clustering in numerous dimensions to achieve useful results in cognitive neuroscience studies.

**Acknowledgments**

This work was supported by the National Natural Science Foundation of China (Grant No. 91748105&81471742) and the Fundamental Research Funds for the Central Universities [DUT2019] in Dalian University of Technology in China.

**References**

Abu-Jamous B, Fa R, Roberts DJ, Nandi AK (2013) Paradigm of Tunable Clustering Using Binarization of Consensus Partition Matrices (Bi-CoPaM) for Gene Discovery Plos One 8 doi:10.1371/journal.pone.0056432

Abu-Jamous B, Fa R, Roberts DJ, Nandi AK (2014) Comprehensive analysis of forty yeast microarray datasets reveals a novel subset of genes (APha-RiB) consistently negatively associated with ribosome biogenesis Bmc Bioinformatics 15 doi:10.1186/1471-2105-15-322

Abu-Jamous B, Fa R, Roberts DJ, Nandi AK (2015) UNCLES: method for the identification of genes differentially consistently co-expressed in a specific subset of datasets Bmc Bioinformatics 16 doi:10.1186/s12859-015-0614-0

Basel Abu-Jamous RF, Asoke K. Nandi (2015) Integrative Cluster Analysis in Bioinformatics Copyright © 2015 John Wiley & Sons, Ltd doi:10.1002/9781118906545

Berg P (2006) Dipole simulator (Version 3.3. 0.4).

Bezdek JC (1981) Pattern recognition with fuzzy objective function algorithms. Pattern recognition with fuzzy objective function algorithms.

Biasiucci A, Franceschiello B, Murray MM (2019) Electroencephalography Curr Biol 29:R80-R85 doi:https://doi.org/10.1016/j.cub.2018.11.052

Brunet D, Murray MM, Michel CM (2011) Spatiotemporal Analysis of Multichannel EEG: CARTOOL Computational Intelligence and Neuroscience doi:10.1155/2011/813870

Charrad M, Ghazzali N, Boiteau V, Niknafs A (2014) Nbclust: An R Package for Determining the Relevant Number of Clusters in a Data Set Journal of Statistical Software 61:1-36

Chen G et al. (2015) Event-related brain potential correlates of prospective memory in symptomatically remitted male patients with schizophrenia Frontiers in Behavioral Neuroscience 9 doi:10.3389/fnbeh.2015.00262

Custo A, Van De Ville D, Wells WM, Tomescu MI, Brunet D, Michel CM (2017) Electroencephalographic Resting-State Networks: Source Localization of Microstates Brain connectivity 7:671-682 doi:10.1089/brain.2016.0476

Delorme A, Makeig S (2004) EEGLAB: an open source toolbox for analysis of single-trial EEG dynamics including independent component analysis Journal of Neuroscience Methods 134:9-21 doi:10.1016/j.jneumeth.2003.10.009

Dinov M, Leech R (2017) Modeling Uncertainties in EEG Microstates: Analysis of Real and Imagined Motor Movements Using Probabilistic Clustering-Driven Training of Probabilistic Neural Networks 11 doi:10.3389/fnhum.2017.00534

Dunn JC (1974) Well-separated clusters and optimal fuzzy partitions Journal of Cybernetics 4:95-104

Fisher RAJM (1921) On the probable error of a coefficient of correlation deduced from a small sample 1:3-32

Fred ALN, Jain AK (2005) Combining multiple clusterings using evidence accumulation Ieee Transactions on Pattern Analysis and Machine Intelligence 27:835-850 doi:10.1109/tpami.2005.113

Goutte C, Toft P, Rostrup E, Nielsen FÅ, Hansen LK (1999) On Clustering fMRI Time Series NeuroImage 9:298-310 doi:https://doi.org/10.1006/nimg.1998.0391

Halkidi M, Vazirgiannis M Clustering validity assessment: Finding the optimal partitioning of a data set. In: Data Mining, 2001. ICDM 2001, Proceedings IEEE International Conference on, 2001. IEEE, pp 187-194

Handy TC e (2009) Brain Signal Analysis: Advances in Neuroelectric and Neuromagnetic Methods MIT Press Scholarship Online:21-53

Jonnalagadda S, Srinivasan R (2009) NIFTI: An evolutionary approach for finding number of clusters in microarray data Bmc Bioinformatics 10 doi:10.1186/1471-2105-10-40

Jung TP, Makeig S, Humphries C, Lee TW, McKeown MJ, Iragui V, Sejnowski TJ (2000) Removing electroencephalographic artifacts by blind source separation Psychophysiology 37:163-178 doi:10.1111/1469-8986.3720163

Kappenman ES, Luck SJJTOhoe-rpc (2012) ERP components: The ups and downs of brainwave recordings:3-30

Karypis G, Kumar V (1998) Multilevelk-way Partitioning Scheme for Irregular Graphs Journal of Parallel and Distributed Computing 48:96-129 doi:https://doi.org/10.1006/jpdc.1997.1404

Kassambara A (2017) Practical Guide to Cluster Analysis in R: Unsupervised Machine Learning vol 1. STHDA,

Kaufman L, Rousseeuw PJ (1987) Clustering by means of medoids. Statistical Data Analysis Based on the L1-Norm and Related Methods. First International Conference.

Kaufman L, Rousseeuw PJ (2009) Finding groups in data: an introduction to cluster analysis vol 344. John Wiley & Sons,

Kawamoto T, Kabashima Y (2017) Cross-validation estimate of the number of clusters in a network Scientific reports 7:3327

Keil A et al. (2014) Committee report: Publication guidelines and recommendations for studies using electroencephalography and magnetoencephalography Psychophysiology 51:1-21 doi:10.1111/psyp.12147

Khanna A, Pascual-Leone A, Michel CM, Farzan F (2015) Microstates in resting-state EEG: Current status and future directions Neuroscience and Biobehavioral Reviews 49:105-113 doi:10.1016/j.neubiorev.2014.12.010

Koenig T, Stein M, Grieder M, Kottlow M (2014) A Tutorial on Data-Driven Methods for Statistically Assessing ERP Topographies Brain Topography 27:72-83 doi:10.1007/s10548-013-0310-1

Kohonen T (1990) THE SELF-ORGANIZING MAP Proceedings of the Ieee 78:1464-1480 doi:10.1109/5.58325

Lehmann D (1989) Microstates of the brain in EEG and ERP mapping studies. In: Brain Dynamics. Springer, pp 72-83

Lehmann D (1990) BRAIN ELECTRIC MICROSTATES AND COGNITION - THE ATOMS OF THOUGHT. Machinery of the Mind: Data, Theory, and Speculations About Higher Brain Function. Birkhauser Boston, Cambridge

Lehmann D, Ozaki H, Pal IJE, neurophysiology c (1987) EEG alpha map series: brain micro-states by space-oriented adaptive segmentation 67:271-288

Lehmann D, Pascual-Marqui RD, Michel C (2009) EEG microstates Scholarpedia 4:7632

Liu C, Abu-Jamous B, Brattico E, Nandi A (2015) Clustering Consistency in Neuroimaging Data Analysis 2015 12th International Conference on Fuzzy Systems and Knowledge Discovery (FSKD):1118-1122

Liu C, Abu-Jamous B, Brattico E, Nandi AK (2017a) Towards Tunable Consensus Clustering for Studying Functional Brain Connectivity During Affective Processing Int J Neural Syst 27:1650042 doi:10.1142/s0129065716500428 %m 27596928

Liu C, Brattico E, Abu-jamous B, Pereira CS, Jacobsen T, Nandi AK (2017b) Effect of Explicit Evaluation on Neural Connectivity Related to Listening to Unfamiliar Music Front Hum Neurosci 11 doi:10.3389/fnhum.2017.00611

Lleti R, Ortiz MC, Sarabia LA, Sanchez MS (2004) Selecting variables for k-means cluster analysis by using a genetic algorithm that optimises the silhouettes Analytica Chimica Acta 515:87-100 doi:10.1016/j.aca.2003.12.020

Luck SJ (2014) An Introduction to the Event-Related Potential Technique, second edition

Mahini R, Cong F (2019) Opt_NC_ERP (Version v1.0.0). doi:http://doi.org/10.5281/zenodo.3345259

Mahini R, Zhou T, Li P, Nandi AK, Li H, Li H, Cong F (2017) Cluster Aggregation for Analyzing Event-Related Potentials. In: Cong F, Leung A, Wei Q (eds) Advances in Neural Networks - ISNN 2017: 14th International Symposium, ISNN 2017, Sapporo, Hakodate, and Muroran, Hokkaido, Japan, June 21–26, 2017, Proceedings, Part II. Springer International Publishing, Cham, pp 507-515. doi:10.1007/978-3-319-59081-3_59

Meila M (2007) Comparing clusterings - an information based distance Journal of Multivariate Analysis 98:873-895 doi:10.1016/j.jmva.2006.11.013

Micah MM, Lucia MD, Brunet D, Michel CM (2009) Principles of Topographic Analyses for Electrical Neuroimaging MIT Press Scholarship Online doi:10.7551/mitpress/9780262013086.003.0002

Michel CM, Brunet D (2019) EEG Source Imaging: A Practical Review of the Analysis Steps Front Neurol 10:18 doi:10.3389/fneur.2019.00325

Michel CM, Koenig T (2017) EEG microstates as a tool for studying the temporal dynamics of whole-brain neuronal networks: A review NeuroImage doi:https://doi.org/10.1016/j.neuroimage.2017.11.062

Michel CM, Koenig T (2018) EEG microstates as a tool for studying the temporal dynamics of whole-brain neuronal networks: A review NeuroImage 180:577-593 doi:https://doi.org/10.1016/j.neuroimage.2017.11.062

Michel CM, Murray MM (2012) Towards the utilization of EEG as a brain imaging tool NeuroImage 61:371-385 doi:https://doi.org/10.1016/j.neuroimage.2011.12.039

Milligan GW, Cooper MC (1985) An examination of procedures for determining the number of clusters in a data set Psychometrika 50:159-179 doi:10.1007/bf02294245

Monti S, Tamayo P, Mesirov J, Golub T (2003) Consensus clustering: A resampling-based method for class discovery and visualization of gene expression microarray data Machine Learning 52:91-118 doi:10.1023/a:1023949509487

Mur A, Dormido R, Duro N, Dormido-Canto S, Vega J (2016) Determination of the optimal number of clusters using a spectral clustering optimization Expert Systems with Applications 65:304-314 doi:10.1016/j.eswa.2016.08.059

Murray MM, Brunet D, Michel CM (2008) Topographic ERP analyses: A step-by-step tutorial review Brain Topography 20:249-264 doi:10.1007/s10548-008-0054-5

Nguyen N, Caruana R Consensus Clusterings. In: Seventh IEEE International Conference on Data Mining (ICDM 2007), 28-31 Oct. 2007 2007. pp 607-612. doi:10.1109/ICDM.2007.73

Oostenveld R, Fries P, Maris E, Schoffelen J-M (2011) FieldTrip: Open Source Software for Advanced Analysis of MEG, EEG, and Invasive Electrophysiological Data Computational Intelligence and Neuroscience 2011:9 doi:10.1155/2011/156869

Pascual-Marqui RD (2002) Standardized low-resolution brain electromagnetic tomography (sLORETA): Technical details Methods Find Exp Clin Pharmacol 24:5-12

Pascualmarqui RD, Michel CM, Lehmann D (1995) SEGMENTATION OF BRAIN ELECTRICAL-ACTIVITY INTO MICROSTATES - MODEL ESTIMATION AND VALIDATION Ieee Transactions on Biomedical Engineering 42:658-665 doi:10.1109/10.391164

Pelleg D, Moore AW X-means: Extending k-means with efficient estimation of the number of clusters. In: Icml, 2000. pp 727-734

Pourtois G, Delplanque S, Michel C, Vuilleumier P (2008) Beyond Conventional Event-related Brain Potential (ERP): Exploring the Time-course of Visual Emotion Processing Using Topographic and Principal Component Analyses Brain Topography 20:265-277 doi:10.1007/s10548-008-0053-6

Rousseeuw PJ (1987) SILHOUETTES - A GRAPHICAL AID TO THE INTERPRETATION AND VALIDATION OF CLUSTER-ANALYSIS Journal of Computational and Applied Mathematics 20:53-65 doi:10.1016/0377-0427(87)90125-7

Ruggeri P, Meziane HB, Koenig T, Brandner C (2019) A fine-grained time course investigation of brain dynamics during conflict monitoring Scientific Reports 9 doi:10.1038/s41598-019-40277-3

Sipola T, Cong F, Ristaniemi T, Alluri V, Toiviainen P, Brattico E, Nandi AK Diffusion map for clustering fMRI spatial maps extracted by independent component analysis. In: Machine Learning for Signal Processing (MLSP), 2013 IEEE International Workshop on, 2013a. IEEE, pp 1-6

Sipola T, Cong F, Ristaniemi T, Alluri V, Toiviainen P, Brattico E, Nandi AK (2013b) DIFFUSION MAP FOR CLUSTERING FMRI SPATIAL MAPS EXTRACTED BY INDEPENDENT COMPONENT ANALYSIS. In: Sanei S, Smaragdis P, Nandi A, Ho ATS, Larsen J (eds) 2013 Ieee International Workshop on Machine Learning for Signal Processing. IEEE International Workshop on Machine Learning for Signal Processing. doi:10.1109/mlsp.2013.6661923

Strehl A, Ghosh J (2002) Cluster ensembles---a knowledge reuse framework for combining multiple partitions Journal of machine learning research 3:583-617

Strehl A, Ghosh J (2003) Cluster ensembles- a knowledge reuse framework for combining multiple partitions Journal of Machine Learning Research 3:583-617 doi:10.1162/153244303321897735

Sugar CA, James GM (2003) Finding the number of clusters in a dataset: An information-theoretic approach Journal of the American Statistical Association 98:750-763 doi:10.1198/016214503000000666

Tan P-N (2006) Introduction to data mining. Pearson Education India,

Tibshirani R, Walther G, Hastie T (2001) Estimating the number of clusters in a data set via the gap statistic J R Stat Soc Ser B-Stat Methodol 63:411-423 doi:10.1111/1467-9868.00293

Topchy AP, Law MHC, Jain AK, Fred AL (2004) Analysis of consensus partition in cluster ensemble. Fourth Ieee International Conference on Data Mining, Proceedings. doi:10.1109/icdm.2004.10100

Tzovara A, Murray MM, Plomp G, Herzog MH, Michel CM, De Lucia M (2012) Decoding stimulus-related information from single-trial EEG responses based on voltage topographies Pattern Recognition 45:2109-2122 doi:https://doi.org/10.1016/j.patcog.2011.04.007

Vega-Pons S, Ruiz-Shulcloper J (2011) A SURVEY OF CLUSTERING ENSEMBLE ALGORITHMS International Journal of Pattern Recognition and Artificial Intelligence 25:337-372 doi:10.1142/s0218001411008683

von Wegner F, Knaut P, Laufs H (2018) EEG Microstate Sequences From Different Clustering Algorithms Are Information-Theoretically Invariant 12 doi:10.3389/fncom.2018.00070

Wackermann J, Lehmann D, Michel CM, Strik WK (1993) Adaptive segmentation of spontaneous EEG map series into spatially defined microstates Int J Psychophysiol 14:269-283 doi:https://doi.org/10.1016/0167-8760(93)90041-M

Yu S, Tranchevent L, Liu X, Glanzel W, Suykens JA, De Moor B, Moreau Y (2012) Optimized data fusion for kernel k-means clustering IEEE Transactions on Pattern Analysis and Machine Intelligence 34:1031-1039